\documentclass{revtex4}
\usepackage{epsfig}
\usepackage{amssymb}
\usepackage{amsmath}
\usepackage{amsfonts}
\usepackage{graphicx}
\usepackage{mathrsfs}
\usepackage[dvips]{color}
\usepackage{multirow}
\newcommand{\fa}{\mathfrak{a}}

\newcommand{\fu}{\mathfrak{u}}

\newcommand{\fn}{{\mathfrak{n}}}

\newcommand{\fz}{\mathfrak{z}}

\newcommand{\bM}{\mathbf{M}}

\newcommand{\cB}{\mathcal{B}}

\newcommand{\cD}{\mathcal{D}}
\newcommand{\cH}{\mathcal{H}}
\newcommand{\cE}{\mathcal{E}}

\newcommand{\cJ}{\mathcal{J}}

\newcommand{\cP}{\mathcal{P}}

\newcommand{\cT}{\mathcal{T}}

\newcommand{\cW}{\mathcal{W}}
\newcommand{\cX}{\mathcal{X}}

\newcommand{\be}{\begin{equation}}
\newcommand{\ee}{\end{equation}}
\newcommand{\bea}{\begin{eqnarray}}
\newcommand{\eea}{\end{eqnarray}}
\newcommand{\nn}{\nonumber}

\newcommand{\ed}{\end{document}}

\newcommand{\bi}{\begin{itemize}}
\newcommand{\ei}{\end{itemize}}

\newcommand{\bce}{\begin{center}}
\newcommand{\ece}{\end{center}}

\newcommand{\sE}{\mathscr{E}}

\begin{document}

\title{Unidirectional Invisibility and $\cP\cT$-Symmetry with Graphene}

\author{Mustafa Sar{\i}saman$^1$}\email{msarisaman@ku.edu.tr}
\affiliation{Department of Physics, Ko\c{c} University, Sar{\i}yer 34450, Istanbul, Turkey}
\author{Murat Tas$^2$}\email{tasm236@gmail.com}
\affiliation{Institute of Nanotechnology, Gebze Technical University, Gebze 41400, Kocaeli, Turkey}

\begin{abstract}

We investigate the reflectionlessness and invisibility properties in the transverse electric (TE) mode solution of a linear homogeneous optical system which comprises the $\mathcal{PT}$-symmetric structures covered by graphene sheets. We derive analytic expressions, indicate roles of each parameter governing optical system with graphene and justify that optimal conditions of these parameters give rise to broadband and wide angle invisibility. Presence of graphene turns out to shift the invisible wavelength range and to reduce the required gain amount considerably, based on its chemical potential and temperature. We substantiate that our results yield broadband reflectionless and invisible configurations for realistic materials of small refractive indices, usually around $\eta = 1$, and of small thickness sizes with graphene sheets of rather small temperatures and chemical potentials. Finally, we demonstrate that pure $\mathcal{PT}$-symmetric graphene yields invisibility at small temperatures and chemical potentials.
\medskip

\noindent {Pacs numbers: 03.65.Nk,  42.25.Bs, 42.60.Da,
24.30.Gd}
\end{abstract}

\maketitle

\section{Introduction}

Emergence of $\cP\cT$-symmetric quantum mechanics~\cite{bender} stunningly elaborates the space of operators yielding real energies into the realm of non-Hermiticity. This abrupt intriguing advance has led to commence studies and applications in various fields of physics~\cite{ijgmmp-2010, bagchi, PT1, PT2, PT4, PT6, PT7, PT8, PT9}, among which $\cP\cT$-symmetry had found the most interest in quantum optics and related fields due to its smoothness to realize experimental investigations and immediate applications~\cite{PT1, PT2, PT10}. A generic $\cP\cT$-symmetric Hamiltonian possesses a potential whose peculiar property is $V(x) = V^{\star} (-x)$~\cite{bender,PT1, PT4}. Complex optical $\cP\cT$-symmetric potentials are realized by the formal equivalence between quantum mechanical Schr\"{o}dinger equation and optical wave equation derived from Maxwell equations. By exploiting optical modulation of the refractive index in the complex dielectric permittivity plane and engineering both optical absorption and amplification, $\cP\cT$-symmetric optical systems can lead to a series of intriguing optical phenomena and devices, such as dynamic power oscillations of light propagation, coherent perfect absorber lasers~\cite{CPA, lastpaper}, spectral singularities~\cite{naimark, p123, pra-2012a, longhi4, longhi3} and unidirectional invisibility~\cite{PT1, PT6, PT7, pra-2017a}.

Advantage of using $\cP\cT$-symmetry in optical systems is that its evolution is measurable through the quantum-optical analogue. In this respect, we investigated the feasibility of realizing unidirectional reflectionlessness and invisibility properties of a $\cP\cT$-symmetric optical slab system by means of a optically active real material using the impressive power of transfer matrix in the framework of quantum scattering formalism in \cite{pra-2017a}. In the present paper we aim to increase the reflectionless and invisible wavelength interval using graphene sheets.

Graphene has a well-documented physical properties and numerous applications which has attracted deep interest and  fashioned scientific limelight for over a decade~\cite{gr1}. Since early discovery of graphene, a vast literature has been emanated and a growing deal of applications has been found~\cite{gr5, gr6, gr7, gr8, naserpour}. The idea that graphene may interact with electromagnetic waves in anomalous and exotic ways, providing new phenomena and applications, has given rise to the study of reflectionlessness and invisibility phenomena in $\cP\cT$-symmetric optical structures with graphenes~\cite{alu2}. Especially recent works on this field performed by the method of two dimensional cloaking and transformation optics fashion up essential motivation of our work~\cite{gr8, naserpour, alu2}, which will use the whole competency of transfer matrix method in scattering formalism~\cite{prl-2009} that exploits spectral singularities and invisibility of electromagnetic fields interacted with an optically active medium~\cite{CPA, lastpaper, p123}.

Invisibility studies in literature bifurcate as cloaking using transformation optics and transfer matrix methods.  The former one exploits the beauty of transformation optics and stunning efficiency of metamaterials~\cite{pendry1}. This approach is based on the fact that object being invisible is to be concealed behind an artificially manufactured material~\cite{metamaterials}. In general, graphene is considered as two dimensional masking material. Another treatment benefits the interferometric methods heading the transfer matrix, which has found a growing interest in recent years~\cite{pra-2012a, longhi4, longhi3, lin1, pra-2015b, pra-2015a, longhi1, pra-2014a, soto, midya, jpa-2014a, pra-2013a, longhi2}, by which we will inquire the role of graphene.

\begin{figure}
    \begin{center}
    \includegraphics[scale=.45]{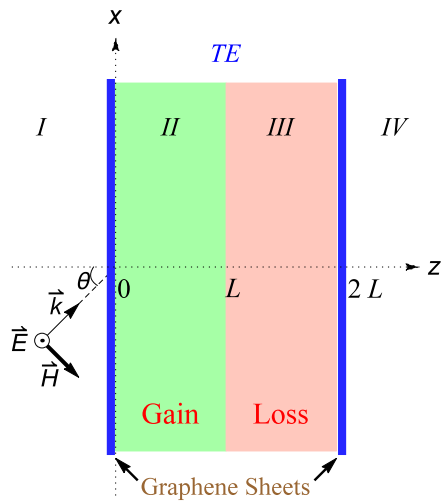}
    \caption{(Color online) Configuration of TE mode of a slab system consisting of a pair of gain and loss layers of thickness $L$ that are covered by graphene sheets in vacuum. The symbols $I$, $I\!I$, $I\! I\! I$ and $I\!V$ respectively label the regions of the space corresponding to $z<0$, $0<z<L$, $L<z<2L$ and $z>2L$. $I\!I$ and $I\!I\!I$ respectively correspond to the gain and loss layers while  $I$ and $I\!V$ represent the vacuum.}
    \label{fig1}
    \end{center}
    \end{figure}

In this paper we conduct a comprehensive study of unidirectional reflectionlessness and invisibility in oblique transverse electric (TE) mode of a $\cP\cT$-symmetric system with graphene to unveil the intriguing traits of transfer matrix as complementary to~\cite{pra-2017a}. Our system is depicted in Fig.~\ref{fig1}.

Our analysis reveals all possible configurations of solutions that support unidirectional reflectionlessness and invisibility. In particular, we obtain analytic expressions for reflectionless and invisible configurations and examine the behaviours of practically most desirable choices of parameters corresponding to the TE waves. We reveal that optimal control of parameters such as gain coefficient, incident angle, slab thickness, temperature and chemical potentials of graphene sheets give rise to a desired outcome of achieving wide wavelength range of unidirectional reflectionlessness and invisibility. Thus, we provide a concrete ground that restrict the gain coefficient, wavelength, slab thickness, incidence angle and temperature and chemical potential of graphene in certain ranges. Optimal values of these parameters should be adjusted in a given system if one desires broadband reflectionless and invisible situations. This provides valuable information about unidirectional reflectionlessness and invisibility for a possible experimental realization of a $\cP\cT$-symmetric slab system with graphene. Out method and hence results are quite reliable for all realistic materials of practical concern.

\section{TE Mode Solution of a Parallel Pair of Slabs with Graphenes}
\label{S2}

Consider a $\cP\cT$-symmetric planar slab system whose exterior surfaces are encompassed by graphene sheets as sketched in Fig.~\ref{fig1}. Suppose that the entire optical system is immersed in air with refractive index $\fn_0 = 1$ and the regions $I\!I$ and $I\!I\!I$ are respectively filled with gain and loss materials having constant complex refractive indices $\fn_1$ and $\fn_2$. Let this system be exposed to an external time harmonic electromagnetic fields, denoted respectively by $\vec \cE$ and $\vec \cH$ as electric and magnetic fields. Maxwell equations describing the interaction of the electromagnetic waves with this system have the form:
    \begin{align}
    &\vec{\nabla}\cdot\vec{\cD} = \rho (z), &&
    \vec{\nabla}\cdot\vec{\cB} = 0,
    \label{equation1}\\
    &\vec{\nabla} \times \vec{\cH}-\partial_t \vec{\cD}=\sigma(z) \vec{\cE}, &&
    \vec{\nabla} \times \vec{\cE} + \partial_t \vec{\cB}=\vec 0,
        \label{equation2}
    \end{align}
where $\vec \cE$ and $\vec \cH$ are connected to $\vec \cD$ and $\vec \cB$ fields via the constitutive relations
    \begin{align*}
    \vec{\cD} := \varepsilon_0 \fz(z)\, \vec{\cE}, &&\vec{\cB}:=\mu_0\vec{\cH},
    \end{align*}
$\varepsilon_0$ and $\mu_0$ are respectively the permeability and permittivity of the vacuum. We defined
    \be
    \fz(z):= \left\{\begin{array}{ccc}
    \fn_1^2 & {\rm for~z}\in I\!I\\
    \fn_2^2 & {\rm for~z}\in I\!I\!I\\
    1 & {\rm otherwise}
    \end{array}\right.
    \label{e1}
    \ee
such that the subindex $j = 1, 2$ represents the gain and loss regions of space respectively as depicted in Fig.~\ref{fig1}. In Maxwell equations (\ref{equation1}) and (\ref{equation2}), $\rho (z)$ and $\sigma (z)$ respectively denote the free charge and conductivity present on the graphene sheets, and therefore expressed as
    \begin{align*}
    \rho (z) := \rho_g^{(1)} \delta(z) + \rho_g^{(2)} \delta(z-2L), &&\sigma (z) := \sigma_g^{(1)} \delta(z) + \sigma_g^{(2)} \delta(z-2L),
    \end{align*}
where $\rho_g^{(j)}$ and $\sigma_g^{(j)}$ are respectively the free charge and conductivity on the $j$-th layer of graphene, with $j = 1, 2$.  Notice that $\rho (z)$ and $\sigma (z)$ are associated to each other by the continuity equation
    \be
    \vec{\nabla}\cdot\vec{\cJ} + \partial_t \rho (z) = 0
    \ee
for the electric current density $\vec{\cJ} := \sigma(z) \vec{\cE}$. The conductivity of graphene sheets has been determined within the random phase approximation in \cite{conductivity-graphene} as the sum of intraband and interband contributions, i.e. $\sigma_{g} = \sigma_{intra}+\sigma_{inter}$, where
    \begin{align}
    \sigma_{intra} &:= \frac{2ie^2k_B T}{\pi\hbar^2\left(\omega + i\Gamma\right)}\ln\left[2\cosh\left(\frac{\mu}{2k_B T}\right)\right], \nn\\
    \sigma_{inter} &:=\frac{e^2}{4\hbar}\left[\frac{1}{2} + \frac{1}{\pi}\arctan\left(\frac{\hbar\omega-2\mu}{2k_B T}\right)-\frac{i}{2\pi}\ln\frac{(\hbar\omega+2\mu)^2}{(\hbar\omega-2\mu)^2 + (2k_B T)^2}\right].
    \label{conductivitydefns}
    \end{align}
Here, $-e$ is the electron charge, $\hbar$ is the reduced Planck's constant, $k_B$ is Boltzmann's constant, $\Gamma$ is the charge carriers scattering rate, $T$ is the temperature, $\mu$ is the chemical potential, and $\hbar\omega$ is the photon energy~\cite{naserpour}. In time harmonic forms, $\vec \cE (\vec{r}, t)$ and $\vec \cH (\vec{r}, t)$ fields are respectively given by $\vec \cE (\vec{r}, t) = e^{-i\omega t} \vec{E}(\vec{r})$ and $\vec \cH (\vec{r}, t)= e^{-i\omega t} \vec{H}(\vec{r})$. Thus, Maxwell equations corresponding to TE wave solutions yield the following form of Helmholtz equation
    \begin{align}
    &\left[\nabla^{2} +k^2\fz(z)\right] \vec{E}(\vec{r}) = 0, &&
    \vec{H}(\vec{r}) = -\frac{i}{k Z_{0}} \vec{\nabla} \times \vec{E}(\vec{r}),
    \label{equation4}
    \end{align}
where $\vec r:=(x,y,z)$, $k:=\omega/c$ is the wavenumber, $c:=1/\sqrt{\mu_{0}\varepsilon_{0}}$ is the speed of light in vacuum, and $Z_{0}:=\sqrt{\mu_{0}/\varepsilon_{0}}$ is the impedance of the vacuum. We stress out that TE waves correspond to the solutions of (\ref{equation4}) for which $\vec \cE(\vec{r})$ is parallel to the surface of the slabs. In our geometrical setup, they are aligned along the $y$-axis. Suppose that in region $I$, incident wave $\vec E(\vec{r})$ adapts a plane wave with wavevector $\vec k$ in the $x$-$z$ plane, specified by
    \begin{align}
    &\vec k=k_x \hat e_x+ k_z \hat e_z, && k_x:=k\sin\theta, &&k_z:=k\cos\theta,
    \end{align}
where $\hat e_x,\hat e_y,$ and $\hat e_z,$ are respectively the unit vectors along the $x$-, $y$- and
$z$-axes, and $\theta\in[-90^\circ,90^\circ]$ is the incidence angle (see Fig.~\ref{fig1}.) Then, the electric field for the TE waves is given by
    \begin{align}
    \vec E(\vec{r})=\sE (z)e^{ik_{x}x}\hat e_y,
    \label{ez1}
    \end{align}
where $\sE$ is solution of the Schr\"odinger equation
    \be
    -\psi^{''}(z)+v(z)\psi(z)=k^2\psi(z),~~~~~~~~~~z\notin\{ 0,L,2L\},
    \label{sch-eq}
    \ee
for the potential $v(z):=k^2[1+\sin^2\theta-\fz(z)]$. The fact that potential $v(z)$ is constant in spaces of interest gives rise to a solution in relevant regions
    \be
    \psi(z):=\left\{\begin{array}{ccc}
    a_0\,e^{ik_z z} + b_0\,e^{-ik_z z} & {\rm for} & z\in I,\\
    a_1\,e^{i{\tilde k}_1z} + b_1\,e^{-i{\tilde k}_1z} & {\rm for} & z\in I\!I,\\
    a_2\, e^{i{\tilde k}_2 z} + b_2\, e^{-i{\tilde k}_2 z} & {\rm for} & z\in I\!I\!I,\\
    a_3\,e^{ik_z z} + b_3\,e^{-ik_z z} & {\rm for} & z\in I\!V,
    \end{array}\right.
    \label{E-theta}
    \ee
where $a_i$ and $b_i$, with $i=0,1,2,3$, are $k$-dependent complex coefficients, and
    \begin{align}
    &{\tilde k}_j:=k\sqrt{\fn^2_j-\sin^2\theta}=k_{z}\tilde\fn_j,
    &&\tilde\fn_j:=\sec\theta \sqrt{\fn^2_j -\sin^2\theta}.
    \label{tilde-parm}
    \end{align}
In particular, $\sE(z)$ is given by the right-hand side of (\ref{E-theta}) with generally different choices for the constants $a_i$ and $b_i$. These coefficients are related to each other by means of appropriate boundary conditions which are expressed by the fact that tangential components of $\vec E$ and $\vec H$ are continuous across the surface while the normal components of $\vec H$ have a step of unbounded surface currents across the interface of graphenes. See Supplemental Material for the secured boundary conditions.

\section{Transfer Matrix Formalism}
\label{S3}

We make use of transfer matrix which is a useful tool in understanding the scattering properties of an optical system. For our two-layer system, transfer matrix can be expressed as the product of transfer matrices of gain and loss regions. If $\bM_{1}$, $\bM_2$ are $2\times 2$ matrices corresponding to the slabs placed in regions of gain and loss respectively, and $\bM=[M_{ij}]$ is the transfer matrix of the composite system, then they all satisfy the composition property $\bM=\bM_2\bM_1$ which is given by
     \begin{align}
    \left[\begin{array}{c}
    a_3\\ b_3\end{array}\right]=\bM \left[\begin{array}{c}
    a_0\\ b_0\end{array}\right].
    \nn
    \end{align}
Hence, the components of transfer matrix are determined to be
    \bea
    &M_{jk} &=\fa_{j}\left[U_{k}(\fu_{j}^{(2)}+j)e^{i\tilde{k}_2 L} + V_{k}(\fu_{j}^{(2)}-j)e^{-i\tilde{k}_2 L}\right],
    \label{M22=x}
    \eea
where subindex $j, k$ denotes $+, -$, and $M_{++}$ represents the $M_{11}$ component and so forth. We also refer to the identifications
    \begin{align}
    &\fa_{\pm} := \frac{e^{\mp2ik_z L}}{8},&&\fu_{\pm}^{(\ell)} := \frac{1\pm \sigma_g^{(\ell)}}{\tilde\fn_{\ell}},\nn\\
    &U_{\pm} := (\tilde{\fn}_1 + \tilde{\fn}_2) (1 \pm \fu_{\pm}^{(1)})e^{i\tilde{k}_1 L} + (\tilde{\fn}_2 - \tilde{\fn}_1) (1 \mp \fu_{\pm}^{(1)})e^{-i\tilde{k}_1 L},\nn\\
    &V_{\pm} := (\tilde{\fn}_2 - \tilde{\fn}_1) (1 \pm \fu_{\pm}^{(1)})e^{i\tilde{k}_1 L} + (\tilde{\fn}_1 + \tilde{\fn}_2) (1 \mp \fu_{\pm}^{(1)})e^{-i\tilde{k}_1 L},
    \label{defns}
    \end{align}
with $\ell = 1, 2$. It is a natural consequence of $\mathcal{PT}$ symmetry that one obtains the following relations
    \begin{align}
    &\fa_{+}\stackrel{\mathcal{PT}}{\longleftrightarrow}\fa_{-},
    &&\tilde{\fn}_1\stackrel{ \mathcal{PT} }{\longleftrightarrow}\tilde{\fn}_2,\nn\\
    &\fu_{\pm}^{(1)}\stackrel{ \mathcal{PT} }{\longleftrightarrow}\fu_{\mp}^{(2)},
    &&\sigma_g^{(1)}\stackrel{ \mathcal{PT} }{\longleftrightarrow}-\sigma_g^{(2)}.
    \label{pt-symmetry-rels}
    \end{align}
Thus, it can be shown that components of transfer matrix (\ref{M22=x}) satisfy the symmetry relations in~\cite{pra-2014c}
 \begin{align}
    &M_{++}\stackrel{\mathcal{PT}}{\longleftrightarrow}M^{*}_{--},
    &&M_{+-}\stackrel{\mathcal{PT}}{\longleftrightarrow}-M^{*}_{+-},
    &&M_{-+}\stackrel{\mathcal{PT}}{\longleftrightarrow}-M^{*}_{-+}.\nn
    \end{align}
We recover that currents flowing on the first graphene sheet is in the opposite direction on the second graphene sheet as a consequence of $\mathcal{PT}$-symmetry. Reflection (left/right) and transmission coefficients are easily constructed to have the relations of the form
   \bea
   &\textrm{R}^{l} &= -\frac{U_{+}(\fu_{-}^{(2)}-1) + V_{+}(\fu_{-}^{(2)}+1)e^{-2i\tilde{k}_2 L}}{U_{-}(\fu_{-}^{(2)}-1) + V_{-}(\fu_{-}^{(2)}+1)e^{-2i\tilde{k}_2 L}},\label{leftreflcoef}\\
   &\textrm{R}^{r} &= \frac{U_{-}(\fu_{+}^{(2)}+1) + V_{-}(\fu_{+}^{(2)}-1)e^{-2i\tilde{k}_2 L}}{U_{-}(\fu_{-}^{(2)}-1) + V_{-}(\fu_{-}^{(2)}+1)e^{-2i\tilde{k}_2 L}}e^{-4ik_z L},\label{rightreflcoef}\\
   &\textrm{T} &=\fa_{-}^{-1}\left[U_{-}(\fu_{-}^{(2)}-1)e^{i\tilde{k}_2 L} + V_{-}(\fu_{-}^{(2)}+1)e^{-i\tilde{k}_2 L}\right]^{-1}.
   \label{transmissioncoef}
   \eea
Information of (right/left) reflection and transmission coefficients give rise to have an insight about unidirectional reflectionlessness and invisibility of the optical system. If the condition $\textrm{R}^{l/r} = 0$ together with $\textrm{R}^{r/l} \neq 0$ is satisfied accordingly, then the optical system is called (left/right) reflectionless. In addition to this condition, one imposes the condition of $\textrm{T}=1$ to reveal  unidirectional invisibility. We examine each case in the following sections.

\section{Unidirectionally Reflectionless and Invisible Potentials}
\label{S4}

The (left/right) reflectionlessness is fulfilled provided that $\textrm{R}^{l/r} = 0$ and $\textrm{R}^{r/l} \neq 0$ hold simultaneously. Thus, one obtains the following conditions for the left/right reflectionlessness
    \begin{align}
    &e^{-2i\tilde{k}_2 L} = \frac{U_{\pm} \left(1 \mp \fu_{\mp}^{(2)}\right)}{V_{\pm} \left(1 \pm \fu_{\mp}^{(2)}\right)},
    &&e^{-2i\tilde{k}_2 L} \neq \frac{U_{\mp} \left(1 \pm \fu_{\pm}^{(2)}\right)}{V_{\mp} \left(1 \mp \fu_{\pm}^{(2)}\right)}.
    \label{unidir-refl}
    \end{align}
It is implied that upper and lower relations are negations of each other if one desires to exhibit the unidirectional reflectionlessness. If both relations are hold simultaneously, the system is said to be bidirectionally reflectionless. See Supplemental Material for the expression of bidirectional reflectionlessness. Besides, uni- or bidirectional invisibility is realized by exposing the condition $\textrm{T}=1$, which yields
    \begin{align}
    e^{-2i\tilde{k}_2 L} = \frac{\fa_{-}^{-1}e^{-i\tilde{k}_2 L} + U_{-} \left(1 - \fu_{-}^{(2)}\right)}{V_{-} \left(1 + \fu_{-}^{(2)}\right)}.
    \label{unittransmis}
    \end{align}
Substitution of (\ref{unittransmis}) in the first expression of (\ref{unidir-refl}) leads to a couple of conditions belonging to the left and right invisible configurations respectively
   \begin{align}
   &e^{\mp 4 i k_z L} = \frac{4\tilde{\fn}_2^2 \left(1-\left[\fu_{\mp}^{(2)}\right]^2\right)}{U_{\pm}V_{\pm}},
   &&e^{\pm 4 i k_z L} \neq \frac{4\tilde{\fn}_2^2 \left(1-\left[\fu_{\pm}^{(2)}\right]^2\right)}{U_{\mp}V_{\mp}}.
   \label{invsibilitycondition1}
   \end{align}
Obviously these conditions point out that unidirectionally invisible configurations are obtained along the curves satisfying foremost relations of (\ref{invsibilitycondition1}) excluding the ranges of the same curves satisfying the second inequalities. Also, See Supplemental Material for the expression of bidirectional invisibility. Only parameters of the optical system involving these relations take part in the invisibility configuration in demand. In principle, equations in (\ref{invsibilitycondition1}) constitute complex expressions in their own rights and split into real and imaginary parts of main expressions. Therefore, it is those real equations that determine the physical parameters of the invisible optical system. Parameters prevalent in our configuration involve the refractive indices of gain and loss regions, thickness $L$ of the slab, incidence angle $\theta$, wavelength $\lambda$ of the shining electromagnetic wave, temperature $T$, charge carrier scattering rate $\Gamma$, and chemical potential $\mu$ of graphene sheets which fulfil the graphene conductivity in (\ref{conductivitydefns}).

\section{Perturbative Approach to Reflectionless and Invisible Potentials: Optimal Conditions of System Parameters}

\begin{figure}
    \begin{center}
    \includegraphics[scale=.50]{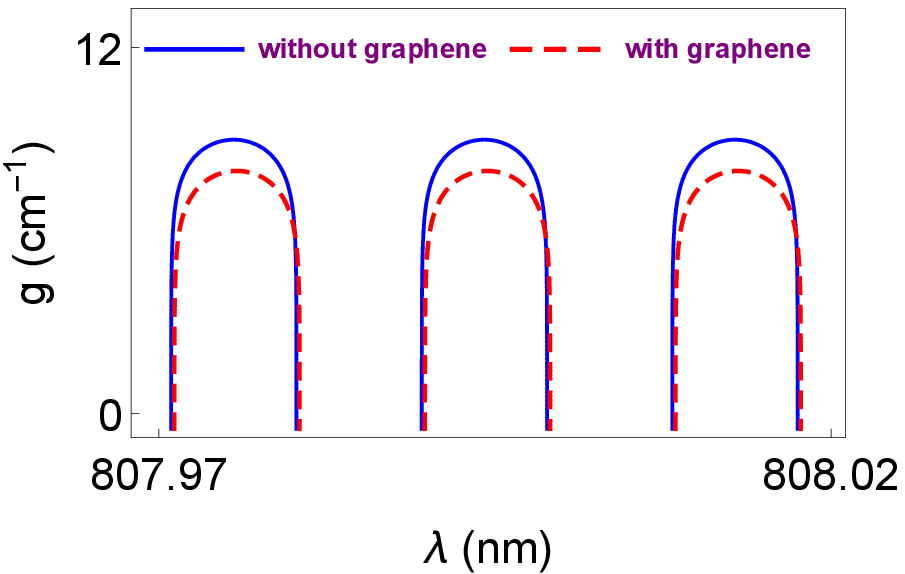}~~~
    \includegraphics[scale=.50]{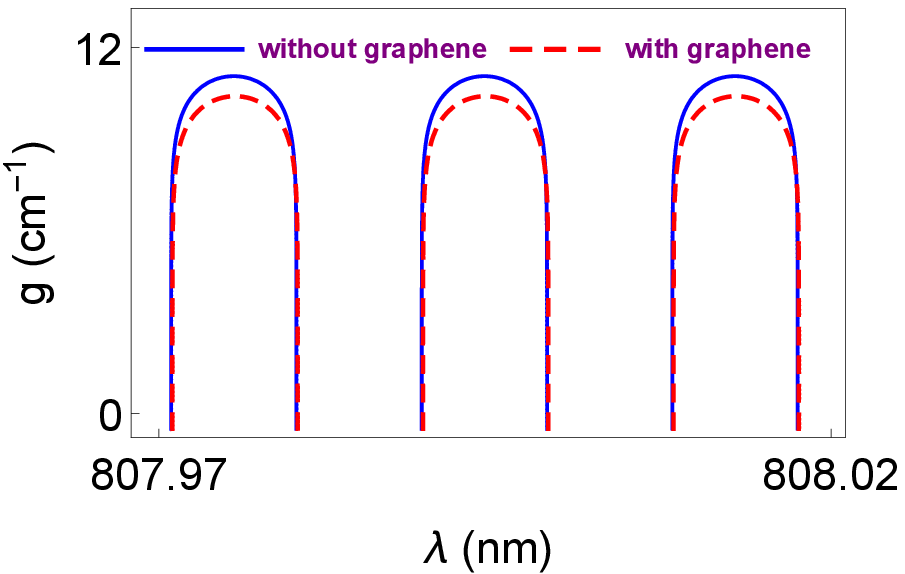}
    \caption{(Color online) The effect of graphene is visualized via gain $g$ versus wavelength $\lambda$ plots for values of bilayer slab in (\ref{specifics}) and graphene features in (\ref{graphenespecifics}). The left and right panels represents left and right zero-reflection conditions respectively. We realize that graphene sheets cause gain values to alter while wavelength patterns shifts along horizontal axis depending on mostly chemical potential of graphene sheets and their temperatures.}
    \label{fig2}
    \end{center}
    \end{figure}

In this section, we derive analytic expressions that reveal general physical consequences of Eq.~\ref{invsibilitycondition1}. Thus we describe (\ref{invsibilitycondition1}) in terms of the quantities of direct physical relevance. We first identify the following
    \begin{align}
    &\tilde\fn:= \tilde\fn_1 =\tilde\fn_2^*,
    &&\sigma_g:= \sigma_g^{(1)}= -\sigma_g^{(2)\ast},
    &&\fu_{\pm}:=\fu_{\pm}^{(1)}= \fu_{\mp}^{(2)\ast}.
    \label{eq251}
    \end{align}
We next denote the real and imaginary parts of $\fn$  by $\eta$ and $\kappa$ so that
    \begin{align}
    &\fn = \eta + i \kappa,
    &&\tilde\fn = \tilde\eta + i \tilde\kappa.
    \label{eq252}
    \end{align}
For most materials of practical concern, one can safely state the condition
   \be
   \left|\kappa\right| \ll \eta - 1 < \eta , \notag
   \ee
such that components $\tilde\eta$ and $\tilde\kappa$ in (\ref{eq252}) give rise to the following respectively, in the leading order of $\kappa$
    \begin{align}
    &\tilde\eta \approx \sec\theta\sqrt{\eta^2-\sin^2\theta},
    &&\tilde\kappa \approx \frac{\sec\theta\,\eta\,\kappa}{\sqrt{\eta^2-\sin^2\theta}}.
    \label{eq253}
    \end{align}
Next, we use another physically applicable parameter, gain coefficient $g$ and recall its definition as given by
   \be
   g:=-2k\kappa = -\frac{4\pi\kappa}{\lambda}.\label{gaincoef}
   \ee
We realize that substitution of Eqs.~\ref{eq251}-\ref{gaincoef} into (\ref{unidir-refl}) and (\ref{invsibilitycondition1}) leads to a couple of real equations connecting physical parameters of our system comprising $\eta$, $g$, $\theta$, $\lambda$, $L$, $T$, $\Gamma$ and $\mu$. The last three parameters are just due to graphene sheets and constribute our constraint relations as long as they do not deform the physical properties of the desired graphene of our use when they are slightly shifted. Parameters $\eta$ and $L$ can be chosen independently such that corresponding $g$ and $\lambda$ can be estimated at a specific incidence angle $\theta$. This is a trustable way of expressing the range of required gain values matching up to wavelength $\lambda$. Therefore, one can adjust the range of wavelength $\lambda$ by playing other independent parameters. We figure out the plenary expression for the required gain values granting unidirectional reflectionlessness as follows. See Supplemental Material for the details.
   \be
   g_{\ell} \approx \frac{\sqrt{\eta^2 -\sin^2\theta}}{\eta L}\,\ln\left[\frac{\mathcal{C}_{\ell} -\sqrt{\mathcal{C}_{\ell}^2 + 4\zeta_{\ell}^{(+)}\zeta_{\ell}^{(-)}}}{2\zeta_{\ell}^{(+)}}\right],\label{perturbgaincoef}
   \ee
where $\zeta_{\ell}^{(\pm)}$ and $ \mathcal{C}_{\ell}$ are denoted by
  \begin{align}
  \zeta_{\ell}^{(\pm)} &:=\left(1 \pm \ell\,\textrm{Re}[\fu_{\ell}]\right)^2 + \left(\textrm{Im}[\fu_{\ell}]\right)^2,\notag\\
  \mathcal{C}_{\ell} &:= -\frac{2\tilde{\eta}}{\tilde{\kappa}}\left\{\left[1-\left(\textrm{Re}[\fu_{\ell}]\right)^2 -\left(\textrm{Im}[\fu_{\ell}]\right)^2\right]\,\sin(2k_z L\tilde{\eta}) +2\ell\, \textrm{Re}[\fu_{\ell}]\,\cos(2k_z L\tilde{\eta})\right\},\notag
  \end{align}
and the index $\ell = + / -$ represents the left/right reflectionless configurations. Thus, bidirectionally reflectionless case occurs at values such that $g_{+} = g_{-}$. To gain a concrete insight into the Eq.~\ref{perturbgaincoef}, we illustrate the revealing physical realizations by means of plots that reflect the basic characteristics of gain coefficient $g$ as a function of wavelength $\lambda$. For this purpose, we lay out Nd:YAG crystals incorporating $\cP\cT$-symmetric bilayer with the following specifications
    \begin{align}
    &\eta = 1.8217, &&L=1~\textrm{cm}, &&\theta = 30^{\circ},\label{specifics}
    \end{align}
and graphene sheets with the characterizations
    \begin{align}
    &T = 300~^{\circ}K, &&\Gamma = 0.1~\textrm{meV}, &&\mu = 0.05~\textrm{eV}.\label{graphenespecifics}
    \end{align}
Fig.~\ref{fig2} signifies the effect of graphene layers on the $\cP\cT$-symmetric bilayer system for the specifications given in (\ref{specifics}) and (\ref{graphenespecifics}). We immediately observe that gain values for unidirectional reflectionlessness are lowered and wavelength range denoted by the size of dome shapes slightly change into the right-hand side. This shows that reflectionless wavelength width and corresponding gain values could be manipulated as desired by setting relevant temperatures and chemical potentials of graphene sheets. In these graphs, we realize that unidirectional reflectionlessness is implemented for certain gain values for the left and right hand-sides as pointed out in \cite{pra-2017a}. Also notice that one requires more gain amounts for right reflectionlessness compared to the left one. See Supplemental Material for the influences of parameters incidence angle, temperature and chemical potential on the reflectionlessness.

We may keep track of the similar approach for the perturbative expressions of the unidirectional invisibility, which give rise to transcendental equations that lead to ultimate solutions among parameters of the optical system. See Supplemental Material for the details and related expressions. Similar to the reflectionless case, the presence of graphene yields the invisible gain value to lower, and the wavelength interval to right-shift  slightly. This means that we can obtain invisible configurations using smaller gain values compared to grapheneless structures by possessing small temperatures at particular extent of chemical potentials. Thus, we can observe invisibility at any wavelength range as we desire. We also observe that one requires higher gain amounts for the right invisibility compared to the left one, and at some wavelengths left and right invisibility gain amounts overlap, which produce bidirectional invisibility.

In case of the finite spread of any monochromatic light, one needs to consider a dispersion effect in refractive index which requires $\kappa_0$ values leading to resonance gain amounts denoted by $g_0$ at resonance wavelength $\lambda_0$. In this case, when the wavelength is spread due to dispersion, the invisible gain amount recedes from $g_0$ to a higher value and wavelength slightly shifts, see \cite{CPA,lastpaper} for the discussion of dispersion.

\section{Exact Analysis of Reflectionlessness and Invisibility}

In view of the perturbative analysis of unidirectional reflectionlessness and invisibility, we are able to illustrate how the parameters of the $\cP\cT$-symmetric slab system surrounded by graphene sheets react to the prescribed phenomena. Optimal values of these parameters serve for the unidirectional reflectionlessness and invisibility properties that we examine. Information about the ranges of system parameters provided by the perturbative analysis can be used to see the extent of uni- or bidirectional reflectionlessness and invisibility. This is a natural consequence and power of transfer matrix formalism that has to be satisfied. Therefore, we investigate the graphs of quantities $\left|R^{l}\right|^2$, $\left|R^{r}\right|^2$ and $\left|T-1\right|^2$ for various and effective system parameters as a level of reflectionlessness and invisibility. For this purpose, we directly use Eqs.~\ref{leftreflcoef}-\ref{transmissioncoef} based on the system parameters we determined perturbatively.

Fig.~\ref{figpi1} clarifies the role of graphene in minimizing the gain value and shifting wavelength range corresponding to reflectionlessness and invisibility phenomena in the $\cP\cT$-symmetric slab system. We situate the slab parameters as in (\ref{specifics}) with different choices of incidence angles, and (\ref{graphenespecifics}) as graphene specifications.  Figures on the left column have incidence angle of $\theta =1.305^{\circ}$ and wavelength $\lambda = 808~\textrm{nm}$ with and without graphenes, and represent prescribed phenomena within $2\%$ of illusion. Upper and lower left figures unveil that whereas there is no left invisibility except for two points in absence of graphene, it occurs till the gain value $g = 2.5~\textrm{cm}^{-1}$ when the graphene sheets are located. Similarly, left and right reflectionless gain values get smaller with the presence of graphene, see Fig.~\ref{figpi1}. For figures on the right, we employ the gain value of $g = 8~\textrm{cm}^{-1}$. Right reflectionlessness and left invisibility in lack of graphene as shown in upper right figure are restricted to the wavelength interval $(808.08~\textrm{nm}, 808.0823~\textrm{nm})$ at incidence angle $\theta = 1.315^{\circ}$. The use of graphene yields right invisibility and left reflectionlessness at incidence angle $\theta = 1.324^{\circ}$. Thus we conclude that graphene results in the left-shifting of wavelength range of reflectionlessness and invisibility, together with right shifting the angle of incidence.
    \begin{figure}
    \begin{center}
    \includegraphics[scale=.45]{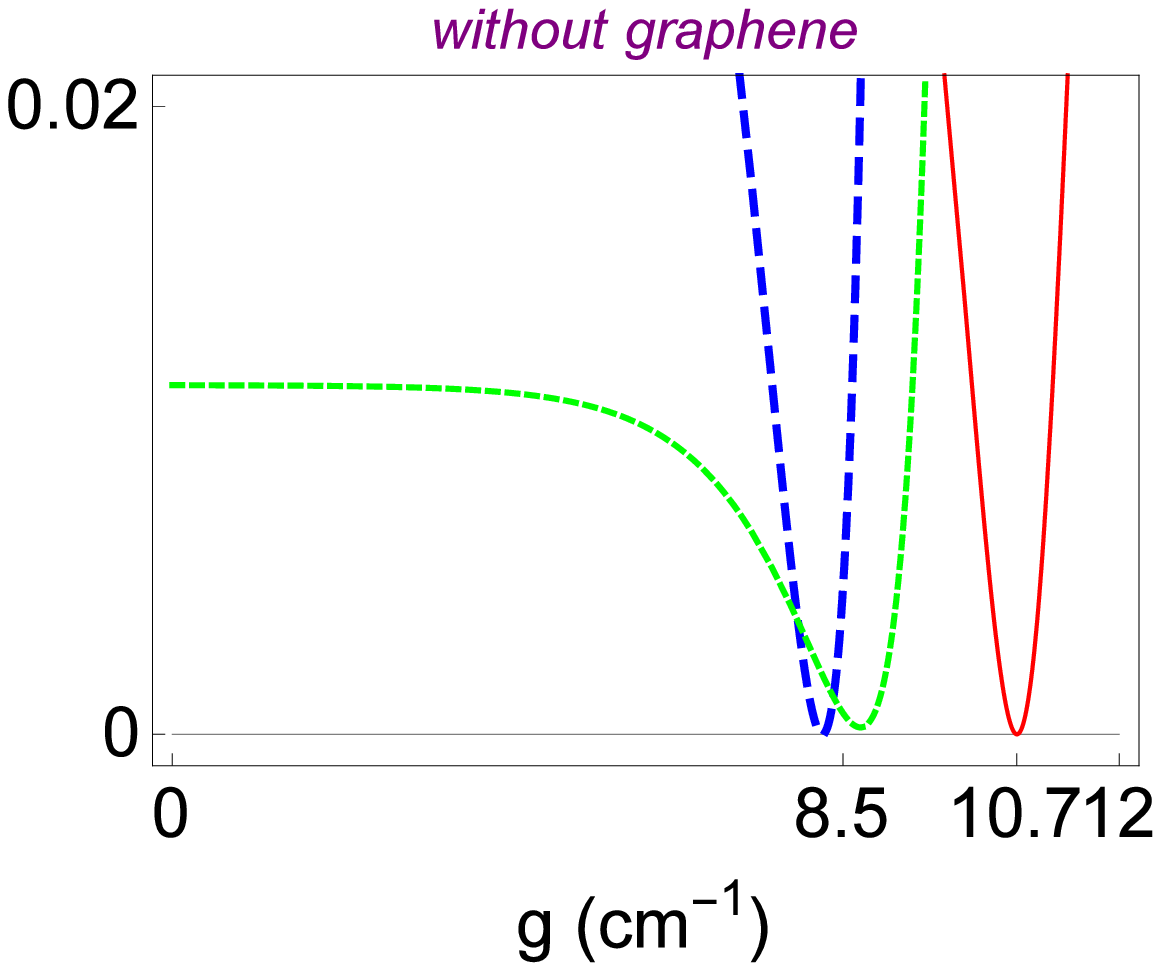}~~~~~
    \includegraphics[scale=.51]{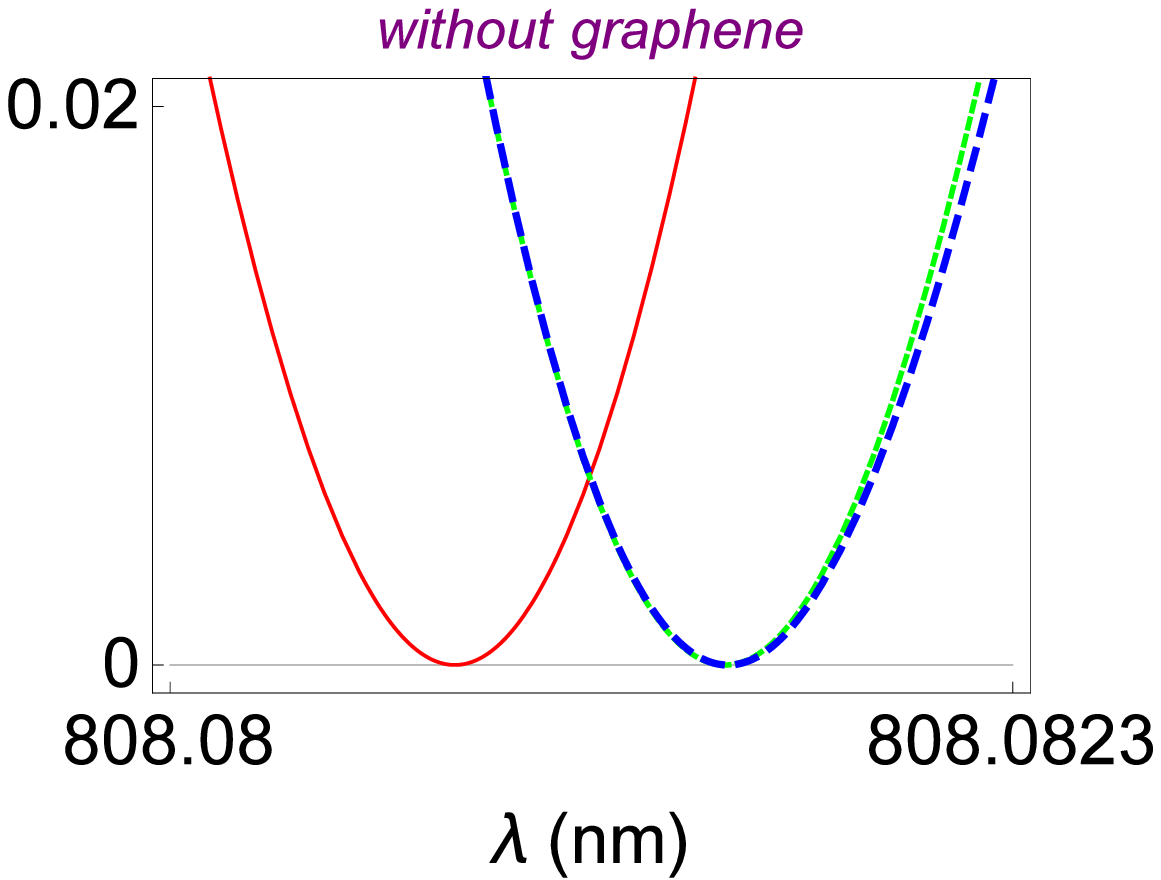}\\
    ~\includegraphics[scale=.45]{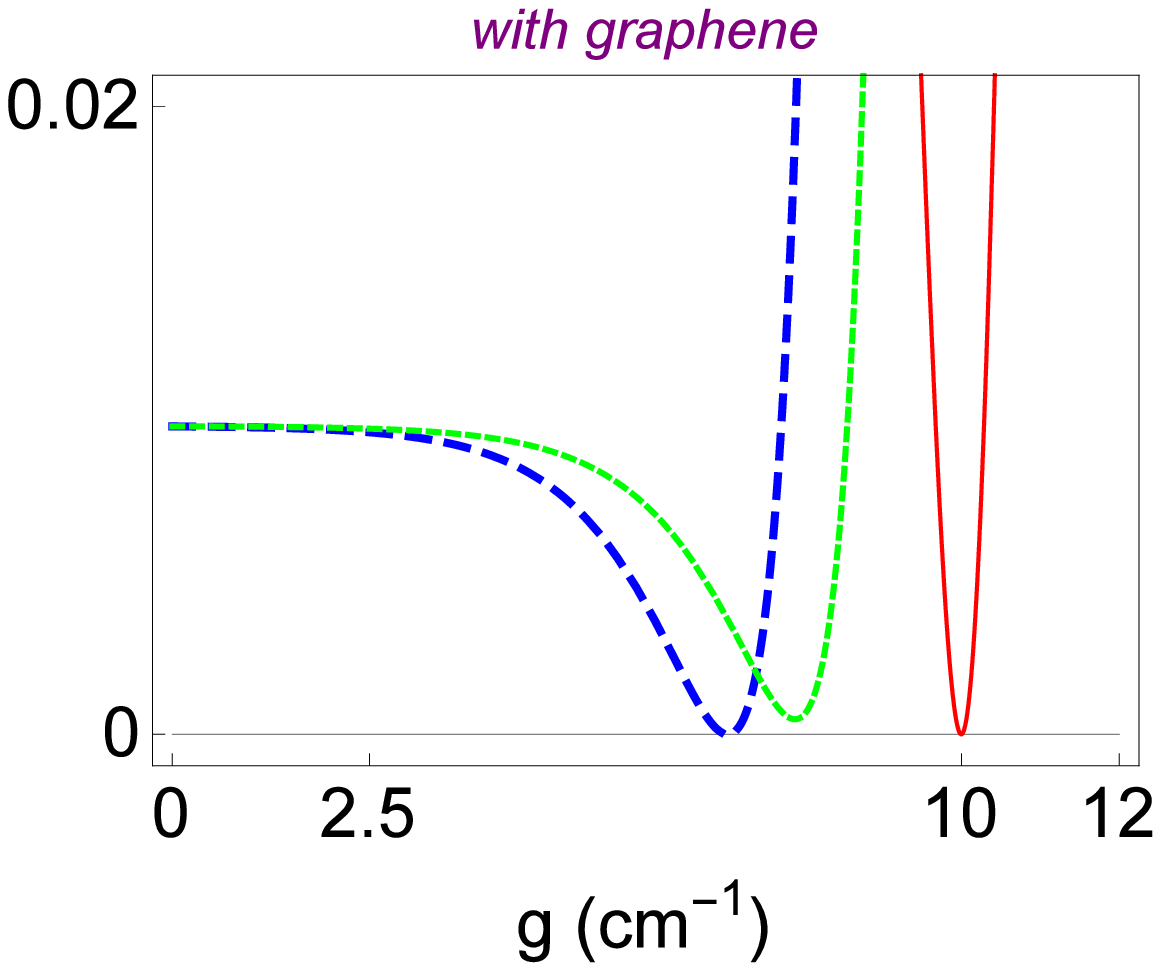}
    \includegraphics[scale=.55]{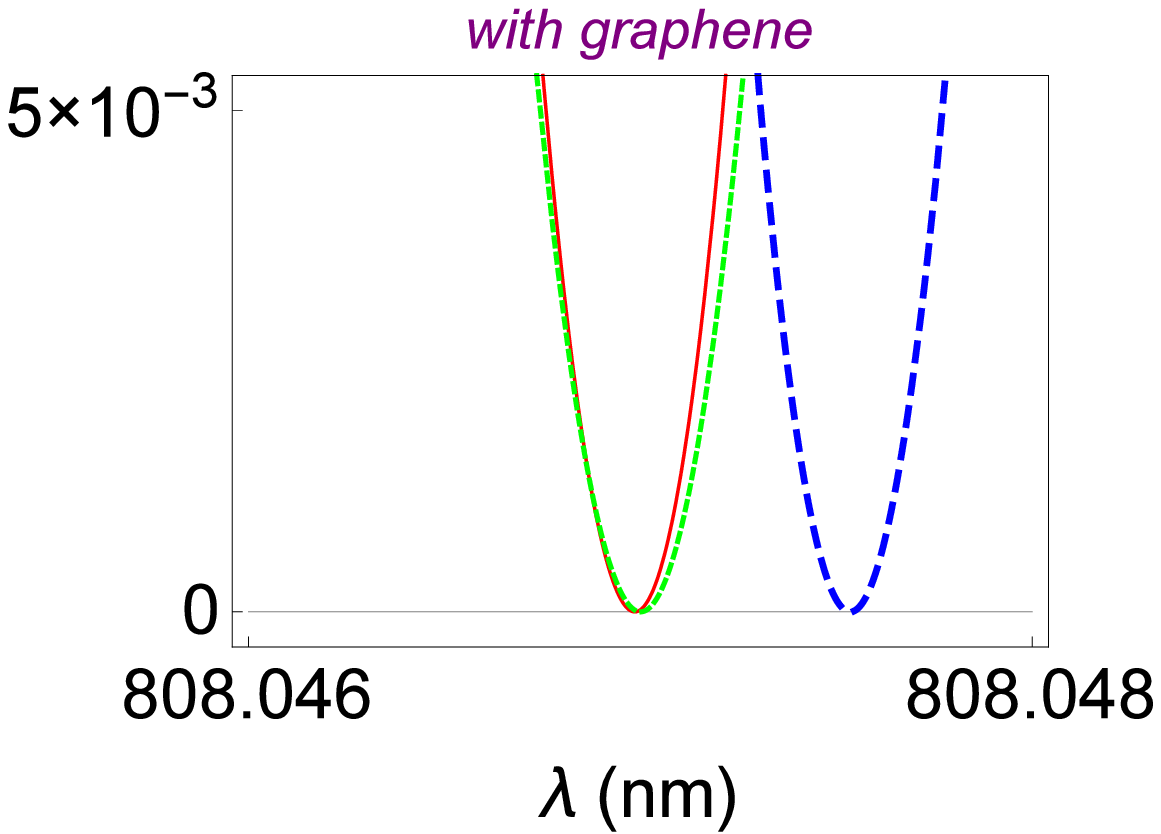}
    \caption{(Color online) Plots of $\left|\textrm{R}^{l}\right|^2$ (thick dashed blue curve), $\left|\textrm{R}^{r}\right|^2$ (thin red curve) and $\left|\textrm{T}-1\right|^2$ (thin dashed green curve) displayed on the vertical axis as a function of gain amount $g$ (left column) and wavelength $\lambda$ (right column) with and without graphenes. It is clearly seen that required gain values for uni- or bidirectional reflectionlessness and invisibility are reduced significantly.}
    \label{figpi1}
    \end{center}
    \end{figure}

See Supplemental Material for the effect of other parameters on quantities $\left|R^{l}\right|^2$, $\left|R^{r}\right|^2$ and $\left|T-1\right|^2$. We observe that graphene results in the left-shifting of wavelength range of reflectionlessness and invisibility, together with right shifting the angle of incidence. Also, it is obvious that all unidirectionally reflectionless and invisible patterns take place at small chemical potentials which are typically less than $\left|\mu\right|=1~\textrm{eV}$ within $1\%$ of precision. Reducing chemical potential very close to zero gives rise to quite reliable patterns. As for the temperature of graphene, we reveal that almost all accessible temperature values produce left and right reflectionless and invisible patterns, but smallest possible temperatures engender the best precise configurations. Angle range of unidirectional reflectionlessness and invisibility decreases, and frequency of invisibility is lower compared to reflectionlessness. The best angle choices are the ones which are very close to zero.

    \begin{figure}
    \begin{center}
    \includegraphics[scale=.45]{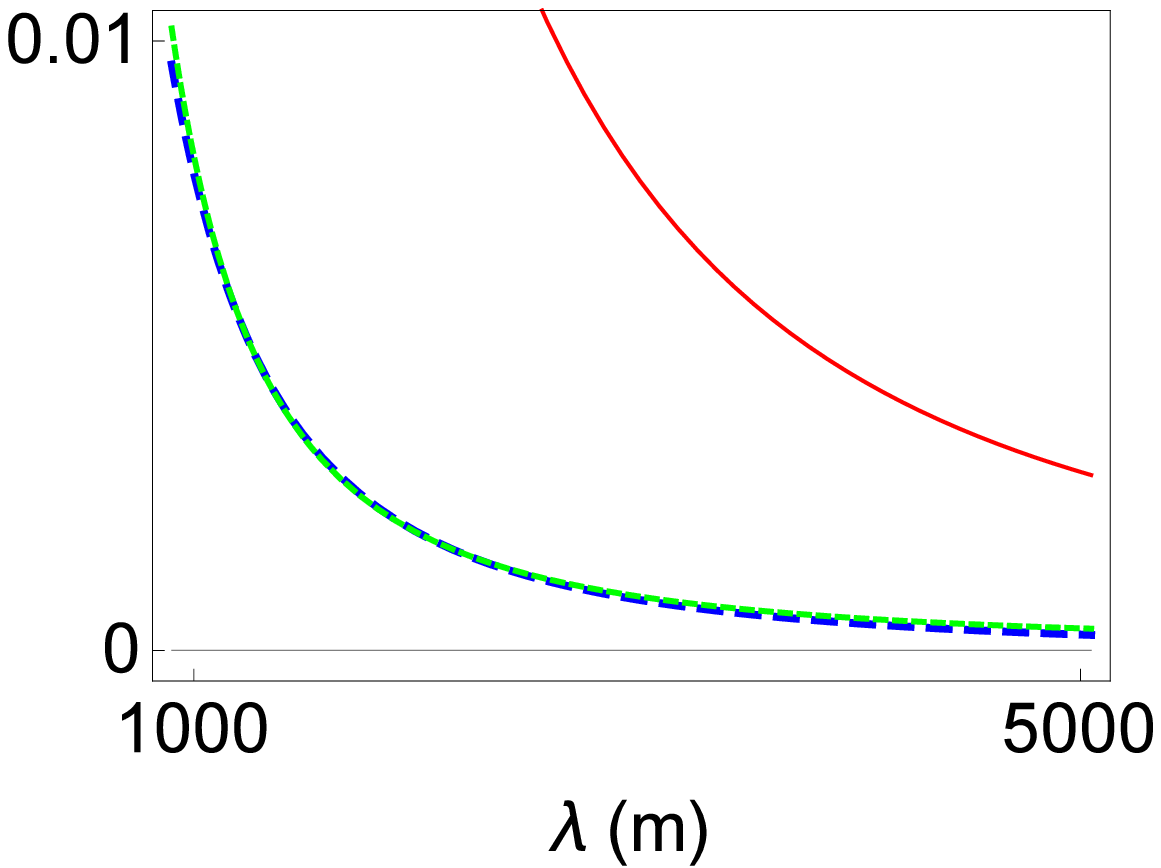}~
    \includegraphics[scale=.45]{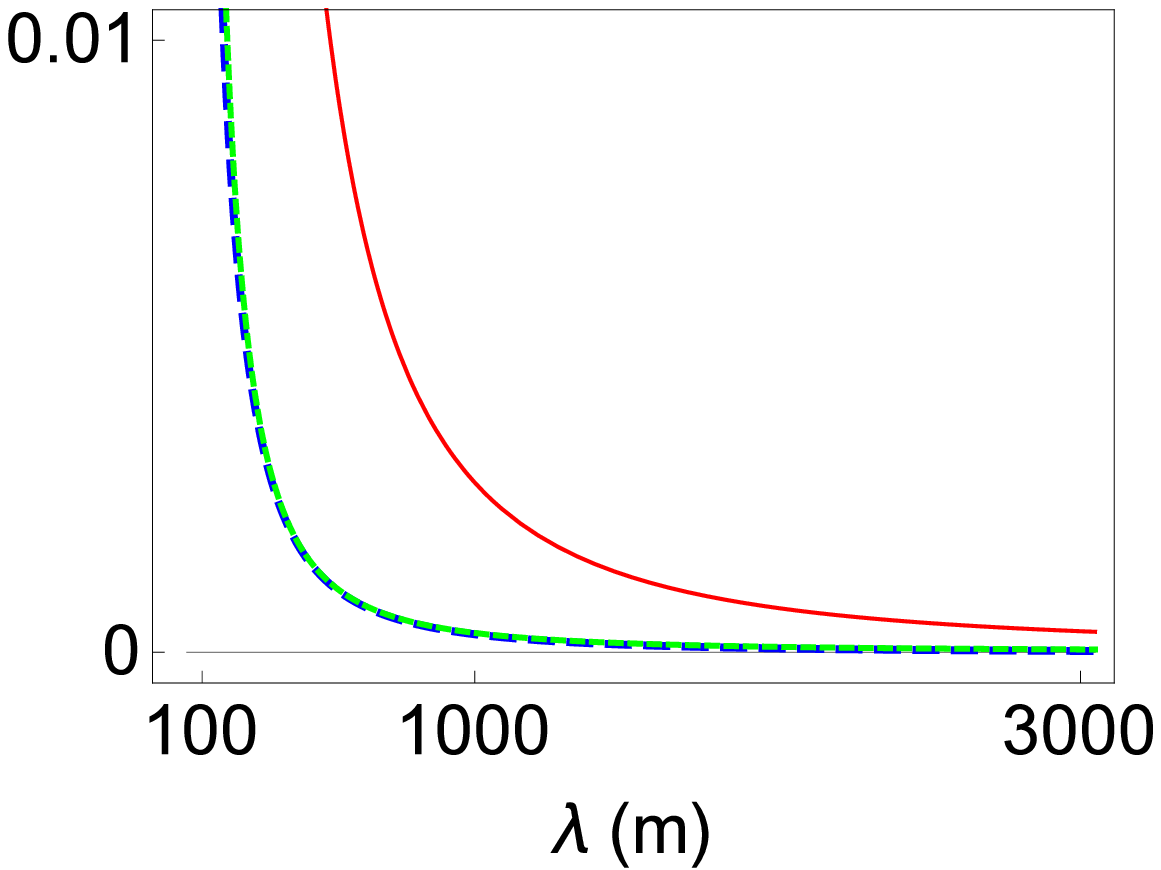}
    \caption{(Color online) Plots of $\left|\textrm{R}^{l}\right|^2$ (thick dashed blue curve), $\left|\textrm{R}^{r}\right|^2$ (thin red curve) and $\left|\textrm{T}-1\right|^2$ (thin dashed green curve) as a function of wavelength $\lambda$ for slab thicknesses $L = 50~\textrm{nm}$ (left figure) and $L = 10~\textrm{nm}$ (right figure) using the slab material as aluminum. As the incidence angle changes, one observes broadband unidirectional reflectionless and invisible configurations.}
    \label{figpi6}
    \end{center}
    \end{figure}
In Fig.~\ref{figpi6}, we reveal the consistency of our findings that parameters of the system has to satisfy in order to generate the desired broadband reflectionless and invisible configurations. For this purpose, we pick out the aluminum as the slab material whose refractive index is $\eta_{Al} = 1.0972$~\cite{alum}, and use two different thickness values as $L=50~\textrm{nm}$ (left figure) and $L=10~\textrm{nm}$ (right figure) to show the effect of thickness. We employ very low temperature and chemical potential values for the graphene as we have explored in Figs.~9 and 10 of Supplemental Material, $T = 5^{\circ}\textrm{K}$ and $\mu = 5\times 10^{-7}~\textrm{eV}$. We realize that distinct phenomena are observed within $1\%$ of precision depending on the incidence angle. First figure on the left gives rise to left invisibility in broad wavelength range $(900~\textrm{nm}, 3300~\textrm{nm})$, and right reflectionlessness in wavelength range $\lambda \geq 2550~\textrm{nm}$ at angle of incidence $\theta = 62.6^{\circ}$. Although we choose a specific incidence angle yielding corresponding phenomena, there is about $1^{\circ}$ angle of extensibility for the same phenomena. As to the right figure with slab thickness $L = 10~\textrm{nm}$, our phenomena in spotlight is observed in a manner that corresponding wavelength range widens and covers a broadband even less than visible spectrum.  When the incidence angle is set to $\theta = 62.8^{\circ}$, we obtain the left invisibility in the wavelength interval $\lambda\geq 165~\textrm{nm}$ and right reflectionlessness in the range $\lambda\geq 500~\textrm{nm}$. See Supplemental Material for the dependence of broadband reflectionless and invisible configurations on the incidence angle. We see that the best invisible and reflectionless configurations are observed once the slab thickness is lowered in nanometer sizes. As it is understood from Fig.~\ref{figpi6}, slab thickness is rather inclusive and crucial in realization of the invisibility phenomenon. In fact, when all other parameters are fixed, slab thickness makes out a behavior of dome patterns that lets invisible gain values in periodically changing intervals of thickness sizes as in Fig.~\ref{fig2}. This shows that not all thickness values are suitable in obtaining invisibility once other parameters are fixed. Thickness tolerance occurs within $1~\textrm{nm}$, and while taking thicknesses around the boundaries of the dome the lowest gain value is obtained. As the size of slab is lowered, the required gain amount rises considerably. In this case one needs a very sensitive slab size corresponding to the dome boundaries, which results in low gain and the widest spectral range for the invisibility.

\section{Pure $\cP\cT$-Symmetric Graphene Invisibility}

\begin{figure}
    \begin{center}
    \includegraphics[scale=.67]{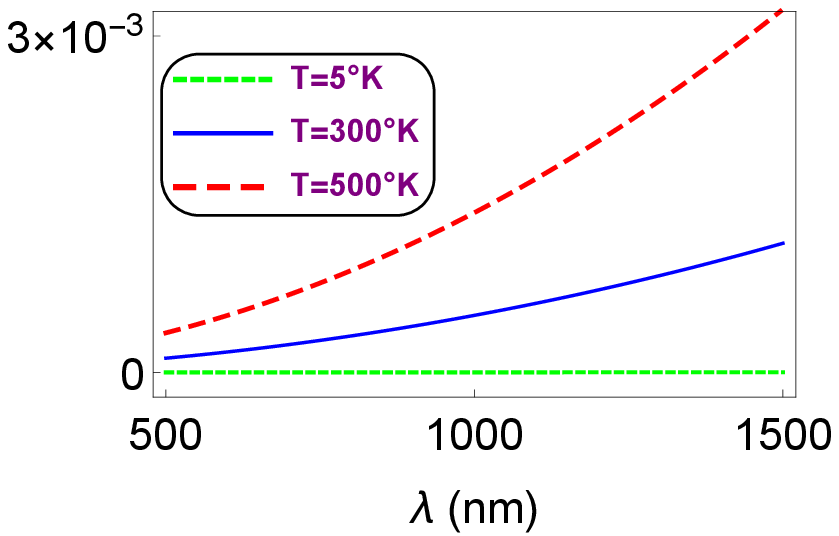}~~~
    \includegraphics[scale=.645]{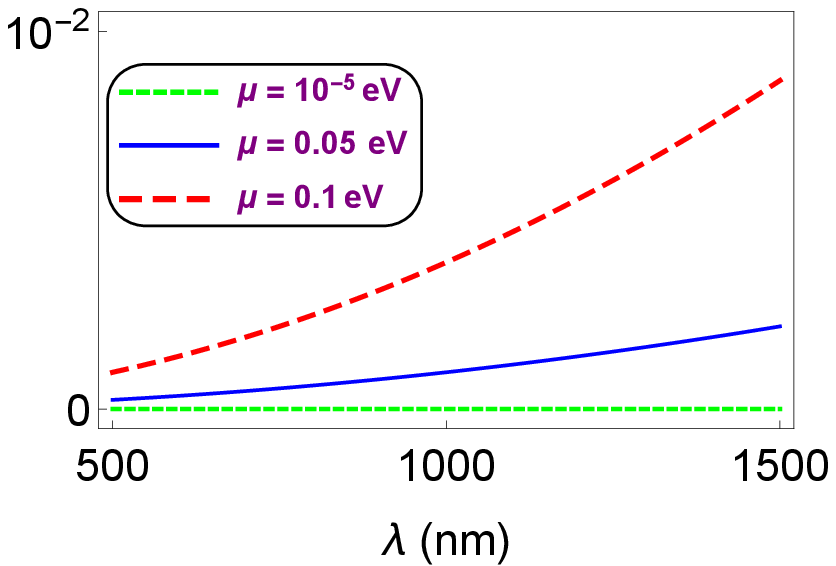}
    \caption{(Color online) Figures showing invisibility of pure $\cP\cT$-symmetric graphene sheets. Vertical axis represents $\left|\textrm{R}^{l}\right|^2 = \left|\textrm{R}^{r}\right|^2 = \left|\textrm{T}-1\right|^2$ and horizontal axis wavelength $\lambda$ for different temperatures $T$ (left panel) and different chemical potentials $\mu$ (right panel). Perfect broadband invisibility is observed at very small temperatures and chemical potentials.}
    \label{figpc2}
    \end{center}
    \end{figure}

Consider a special case that our $\cP\cT$-symmetric slab is removed by setting slab thickness $L$ to zero, i.e. $L = 0$. This case in fact contains a slab material content with zero thickness between graphene layers, namely there is an implicit interface ingredient between graphene sheets. One requires also $\fn_1 = \fn_2 = 0$ to suppress this interface substance. But as we see below, even this case does not affect the graphene invisibility feature since reflection and transmission amplitudes do not depend upon refractive indices $\fn_1$ and $\fn_2$. Setting $L=0$ simply yields that
  \begin{align}
  &\textrm{R}^{l} = \textrm{R}^{r} = \frac{\textrm{Im}[\sigma_g]}{\textrm{Im}[\sigma_g] + i}, && \textrm{T} = \frac{i}{\textrm{Im}[\sigma_g] + i}. \notag
  \end{align}
Hence, notice that $\textrm{R}^{l} = \textrm{R}^{r} = 1-\textrm{T}$ and $\left|\textrm{R}^{l}\right|^2 = \left|\textrm{R}^{r}\right|^2 = \left|\textrm{T}-1\right|^2$. Therefore, at all temperatures and chemical potentials one observes bidirectional invisibility for pure $\cP\cT$-symmetric graphene sheets. The measure of invisibility is improved by reducing temperature $T$ and chemical potential $\mu$ as sketched in diagrams of Fig.~\ref{figpc2}. For the left figure we take $\mu = 10^{-5}~\textrm{eV}$ for the chemical potential and for the right figure $T = 5~^{\circ}K$ for the temperature. We observe that even for large temperatures and chemical potentials, invisibility level is less than $1\%$.  Perfect broadband invisibility is realized at near absolute zero temperature and zero chemical potential.

\section{Concluding Remarks}
\label{S9}

This paper aims at engendering the role played by graphene sheets over the unidirectional reflectionlessness and invisibility, and their optical realizations. We provide that graphene sheets covering the $\cP\cT$-symmetric slab system respect the trait of $\cP\cT$-symmetry. We showed that overall $\cP\cT$-symmetry condition causes surface currents flowing in opposite directions. $\cP\cT$-symmetry guarentees the existence of reflectionlessness and invisibility, due to its power to control the system parameters as distinct from non-$\cP\cT$-symmetric structures, once the appropriate parameters are inserted into the optical system whereas just one gain or loss regions do not yield that phenomenon.

We employ the elegance of transfer matrix method to extract information about reflectionless and invisible configurations, which emphasizes the power of boundary conditions arising from the solutions directly coming from Maxwell equations. The availability of graphene appears in boundary conditions in the form of complex function $\fu_{\pm}^{(j)}$, see Eq.~\ref{defns}. In our analysis, we are able to derive the exact uni- or bidirectional expressions as seen in Eqs.~\ref{unidir-refl}-\ref{invsibilitycondition1} relating parameters of the $\cP\cT$-symmetric optical system with graphene. We utilized the perturbative approach as a tool to demonstrate the optimal conditions arising from system parameters.

In view of our findings, one can shape reflectionless and invisible patterns provided that parameters specifying the graphene and bilayer slab system are well-adjusted. Our primary purpose to place graphene was to widen reflectionless and invisible wavelength range together with the wide incidence angle interval $\Delta\theta$. We explore that this is achieved at small slab thicknesses, which is usually scaled down to nanometer levels, small temperatures, at which best results are obtained near absolute zero value, and quite small chemical potentials, which are typically ones around zero values. Depending upon incidence angle, left or right reflectionless and invisible configurations can be arranged. Furthermore, it is observed that graphene causes the necessary gain amount to lower, the wavelength interval defined for desired phenomena to shift. Another natural consequence is that incidence angles yielding reflectionless and invisible patterns dislocate in compatible with the chemical potentials and temperatures of graphene sheets.

We find out that broadband invisibility is realized at very small thickness values of slab which is less than wavelength $\lambda$, which is typically around nanometer scales, together with a corresponding material type in the slab whose refractive index is small, characterictically around $\eta = 1$ and $\eta < 1.5$, as seen in Fig.~\ref{figpi6}. This is because the effect of graphene is increased when the amount of slab is decreased. In light of this consideration, we are able to show that a pure $\cP\cT$-symmetric graphene structure builds a perfect invisibility at tiny temperatures and chemical potentials near zero. Finally, since broadband reflectionless and invisible configurations are very sensitive to material type which is denoted by refractive index, an extensive analysis of different kinds of material types could be  considered through our method that will feature the usage of metamaterials in $\cP\cT$-symmetric graphene structures.\\[6pt]

\newpage

\section*{Supplemental Material}

Table~\ref{table01} displays the corresponding set of boundary conditions.  \begin{table}[!htbp]
    \begin{center}
	{%\small
    \begin{tabular}{|c|c|}
    \hline
    &\\[-10pt]
    $z=0$ &
    $\begin{aligned}
    & a_1+ b_1 = a_0 + b_0 , && a_1 - b_1 = \fu_{+}^{(1)} a_0- \fu_{-}^{(1)} b_0\\[3pt]
    \end{aligned}$\\
    \hline
    &\\[-10pt]
    $z=L$ & $\begin{aligned}
    & a_1 e^{i{\tilde k}_1 L} + b_1 e^{-i{\tilde k}_1 L}= a_2 e^{i{\tilde k}_2 L} + b_2 e^{-i{\tilde k}_2 L} \\[3pt]
    & \tilde\fn_1 (a_1 e^{i{\tilde k}_1 L} - b_1 e^{-i{\tilde k}_1 L})=
    \tilde\fn_2(a_2 e^{i{\tilde k}_2 L} - b_2 e^{-i{\tilde k}_2 L})
    \end{aligned}$\\[-8pt]
    &\\
    \hline
     &\\[-10pt]
     $z=2L$ & $\begin{aligned}
    & a_2 e^{2i{\tilde k}_2 L} + b_2 e^{-2i{\tilde k}_2 L}= a_3 e^{2ik_{z}L} + b_3 e^{-2ik_{z}L} \\[3pt]
    & a_2 e^{2i{\tilde k}_2 L} - b_2 e^{-2i{\tilde k}_2 L}=
   \fu_{-}^{(2)} a_3 e^{2ik_{z}L} - \fu_{+}^{(2)} b_3 e^{-2ik_{z}L}
    \end{aligned}$\\[-8pt]
    &\\
    \hline
    \end{tabular}}
    \vspace{6pt}
    \caption{Boundary conditions for TE waves. Here $\fu_{\pm}^{(\ell)}$ are defined by (\ref{u=}).}
    \label{table01}
    \end{center}
\end{table}

Here the quantities $\fu_{\pm}^{(\ell)}$ involved are identified by
     \be
     \fu_{\pm}^{(\ell)} := \frac{1\pm \sigma_g^{(\ell)}}{\tilde\fn_{\ell}}.
     \label{u=}
     \ee

Bidirectional reflectionlessness arises provided that
    \be
    \frac{V_{-}\,U_{+}}{V_{+}\,U_{-}} = \frac{\left(1 + \fu_{+}^{(2)}\right)\,\left(1 + \fu_{-}^{(2)}\right)}{\left(1 - \fu_{+}^{(2)}\right)\,\left(1 - \fu_{-}^{(2)}\right)}. \label{bidirectionalreflectionless}
    \ee

Bidirectional invisibility occurs when the following condition holds
    \be
    U_{+}\,U_{-}\,V_{+}\,V_{-} = 16\,\tilde{\fn}_2^4\,\left(1-\left[\fu_{+}^{(2)}\right]^2\right)\,\left(1-\left[\fu_{-}^{(2)}\right]^2\right).\label{bidirectionalinvisibility}
    \ee

To exploit the physical consequence of (17) for the unidirectional reflectionlessness in the main text, we make use of Eqs.~20 - 23 in (17) to split into the real and imaginary parts. Fortunately, real and imaginary parts in the leading order of $\kappa$ give rise to the same equation which is expressed by
   \be
   \zeta_{\ell}^{(+)}\,e^{\tilde{g}L} - \zeta_{\ell}^{(-)}\,e^{-\tilde{g}L} \approx \mathcal{C}_{\ell},\label{pertreflectnlessness}
   \ee
where we denote the following
  \begin{align}
  \zeta_{\ell}^{(\pm)} &:=\left(1 \pm \ell\,\textrm{Re}[\fu_{\ell}]\right)^2 + \left(\textrm{Im}[\fu_{\ell}]\right)^2,~~~~~~~~~~ \tilde{g} := \frac{\eta g}{\sqrt{\eta^2 - \sin^2\theta}} ,\notag\\
  \mathcal{C}_{\ell} &:= -\frac{2\tilde{\eta}}{\tilde{\kappa}}\left\{\left[1-\left(\textrm{Re}[\fu_{\ell}]\right)^2 -\left(\textrm{Im}[\fu_{\ell}]\right)^2\right]\,\sin(2k_z L\tilde{\eta}) +2\ell\, \textrm{Re}[\fu_{\ell}]\,\cos(2k_z L\tilde{\eta})\right\},\notag
  \end{align}
and the index $\ell = + / -$ represents the left/right reflectionless configurations. The symbol $\approx$ implies that we consider terms up to the order of $\kappa$. But, it is easy to show that perturbative approximation we realized coincides with the exact expressions in the realm of physical situations we consider. Thus, Eq.~24 in the main text leads to the perturbative expressions for the gain coefficients corresponding to the left/right reflectionlessness
   \be
   g_{\ell} \approx \frac{\sqrt{\eta^2 -\sin^2\theta}}{\eta L}\,\ln\left[\frac{\mathcal{C}_{\ell} -\sqrt{\mathcal{C}_{\ell}^2 + 4\zeta_{\ell}^{(+)}\zeta_{\ell}^{(-)}}}{2\zeta_{\ell}^{(+)}}\right].\label{perturbgaincoef}
   \ee

\begin{figure}
    \begin{center}
    \includegraphics[scale=.60]{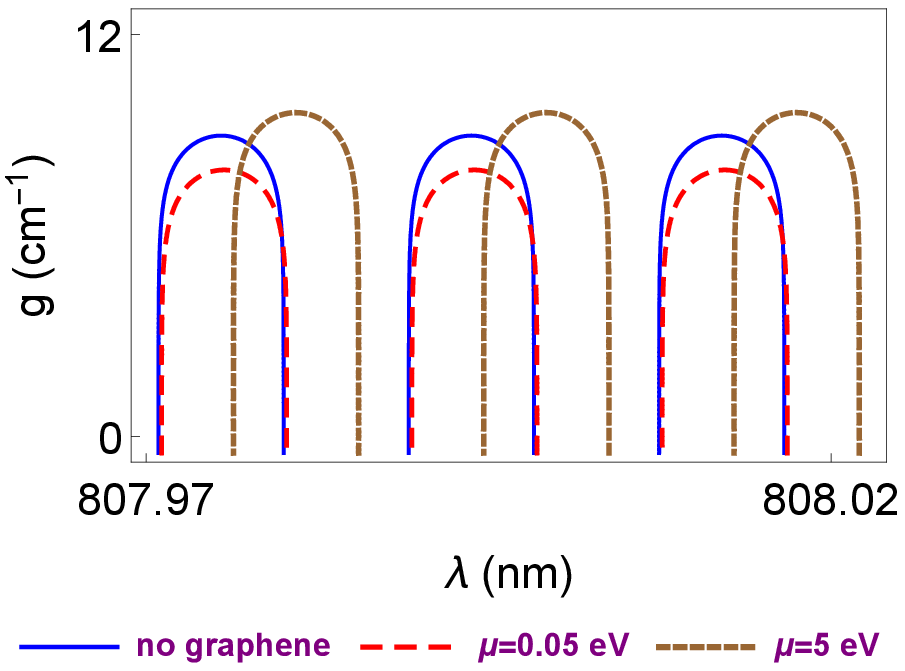}~~~
    \includegraphics[scale=.60]{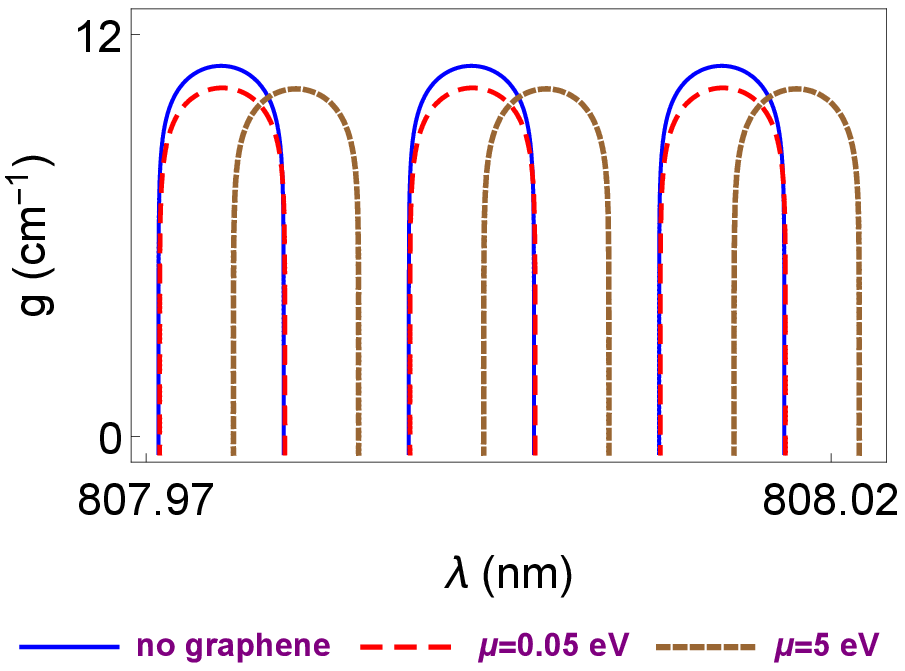}
    \caption{(Color online) Plots of gain $g$ as a function of wavelength $\lambda$ for various chemical potential values of graphene sheets. Left and right boards display the left and right reflectionlessness situations. It is clear that a substantial modulation of chemical potential leads the wavelength range to move significantly whereas the corresponding gain values wiggle up and down.}
    \label{fig3}
    \end{center}
    \end{figure}

One picks out that the presence of graphene is mostly distinguished by the effect of chemical potential as can be seen in Fig.~\ref{fig3} explicitly. We realize that when the chemical potential is increased dramatically, wavelength patterns evolve correspondingly. Besides, necessary gain amounts fluctuate in the vertical direction.

\begin{figure}
    \begin{center}
    \includegraphics[scale=.65]{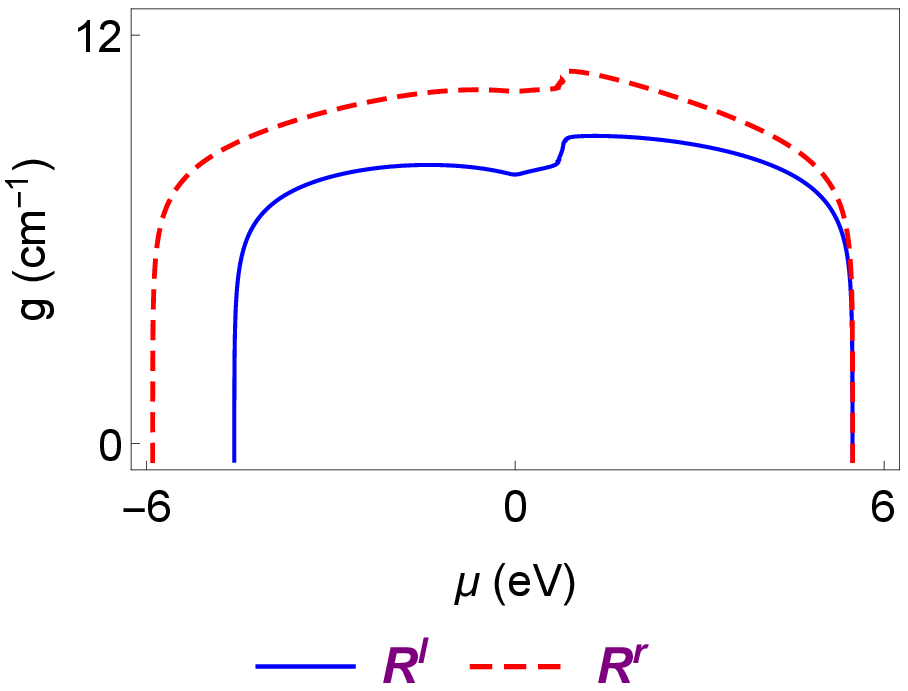}~~~
    \includegraphics[scale=.72]{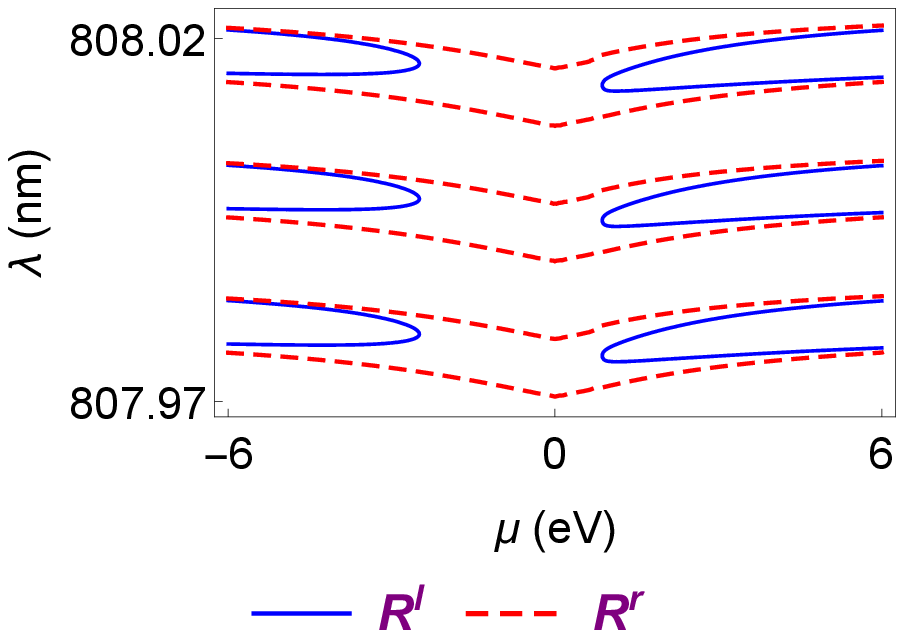}
    \caption{(Color online) Plots showing the dependence of gain $g$ (left panel) and wavelength $\lambda$ (right panel) on the chemical potential $\mu$. Right and left reflection quantities are displayed by solid and dashed curves respectively.}
    \label{fig4}
    \end{center}
    \end{figure}

In Fig.~\ref{fig4}, one can observe the involvement of chemical potential $\mu$ on quantities gain $g$ (on the left side) and wavelength $\lambda$ (on the right side). For the left figure, we set wavelength to $\lambda = 808~\textrm{nm}$ together with the other parameters' values in (25) and (26) in the main text. Moreover, the gain amount is set to $g = 10~\textrm{cm}^{-1}$ with the same parameter values on the right figure. We observe that subject to the wavelength employed there is a maximum limit of chemical potential after which no gain value could be attained in order to obtain a unidirectional reflectionlessness. By adjusting chemical potential at the specified portions, it is tenable to get as small gain amount as possible. Also notice that one needs more gain values for the right reflectionlessness as compared to the left one. On the other hand, for smaller chemical potential, no wavelength value can produce left reflectionlessness. Once the chemical potential is increased to a sufficient extend, left and right reflectionlessness is observed at the same wavelengths. Wavelength configurations indicate that there is an almost periodically repeated patterns such that only certain wavelength widths give rise to unidirectional reflectionlessness.

\begin{figure}
    \begin{center}
    \includegraphics[scale=.65]{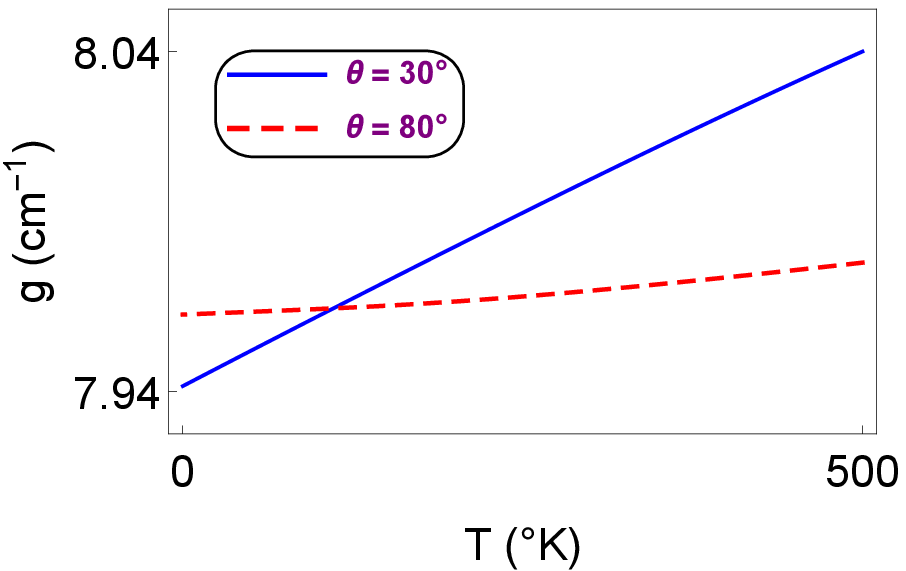}~~~
    \includegraphics[scale=.65]{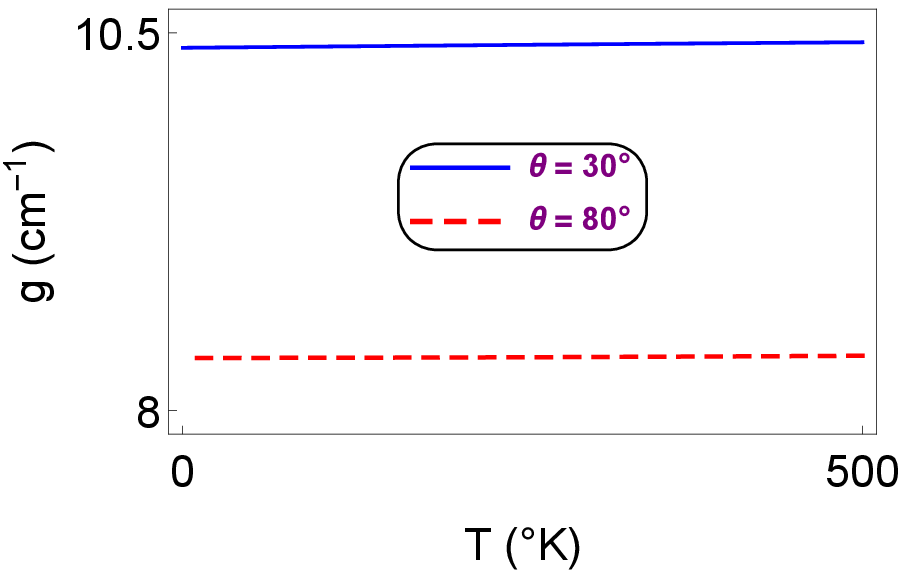}
    \caption{(Color online) Plots of gain value $g$ as a function of temperature $T$ for various incidence angles. The left/right panel represents left/right reflectionless configurations. Temperature change results in a slight change in gain. Left reflectionlessness reacts to the incidence angle variations considerably. }
    \label{fig5}
    \end{center}
    \end{figure}

In Fig.~\ref{fig5}, one observes how the gain amounts depend upon the graphene temperature $T$ for different incidence angles. We set wavelengths to $\lambda = 808.012~\textrm{nm}$ and use system parameters (25) and (26) in the main text. Left/right figure belongs to the left/right reflectionlessness. Notice that at fixed incidence angle, reaction of gain amount on the moderate temperate changes is almost stable, and it increases slightly with the rise of temperature. Once the incidence angle changes, the right reflectionlessness responds acutely considering the left one.

\begin{figure}
    \begin{center}
    \includegraphics[scale=.60]{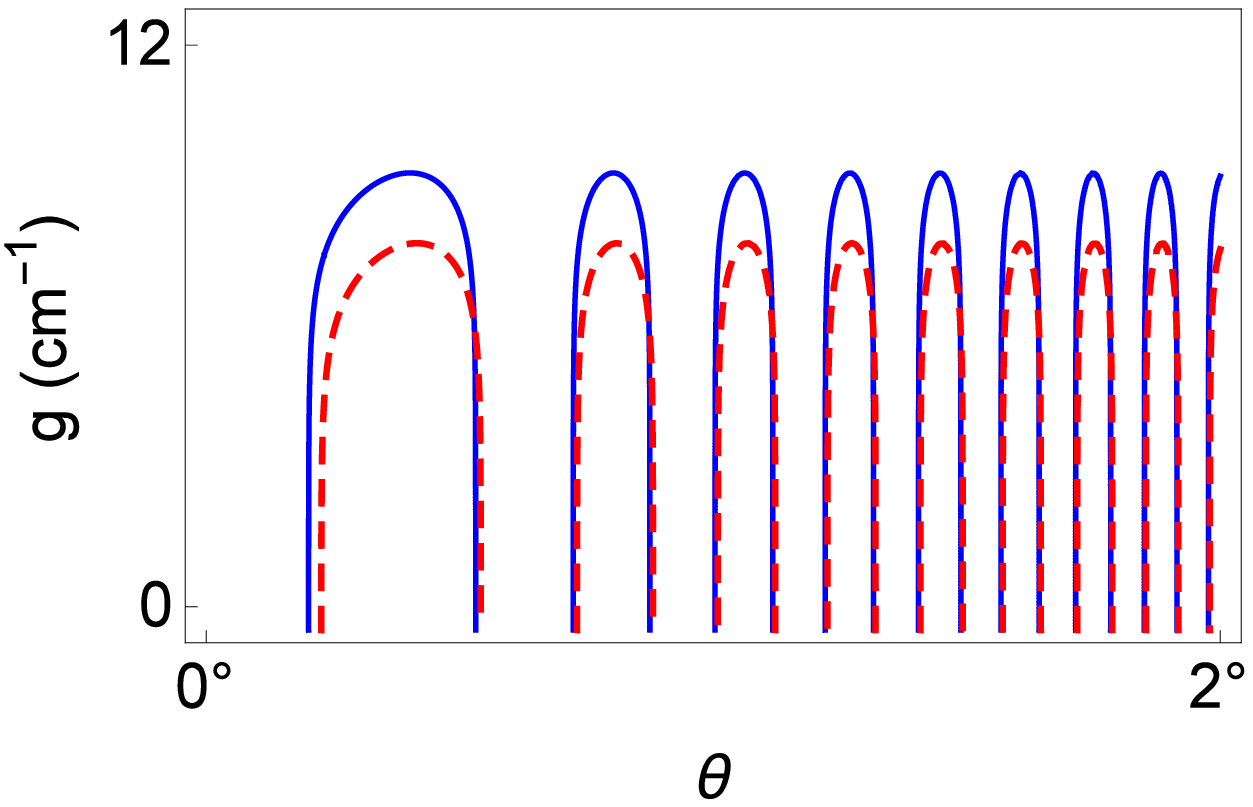}~~~
    \includegraphics[scale=.60]{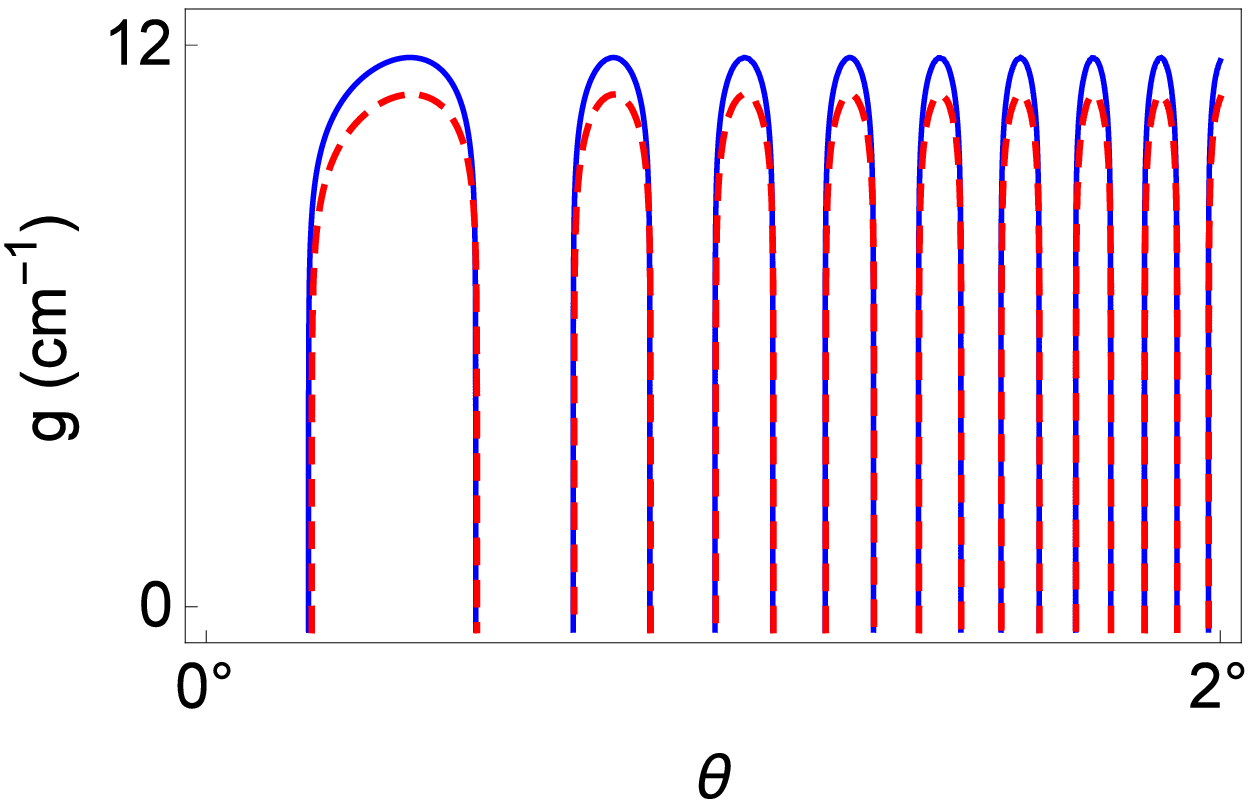}\\
    \includegraphics[scale=.60]{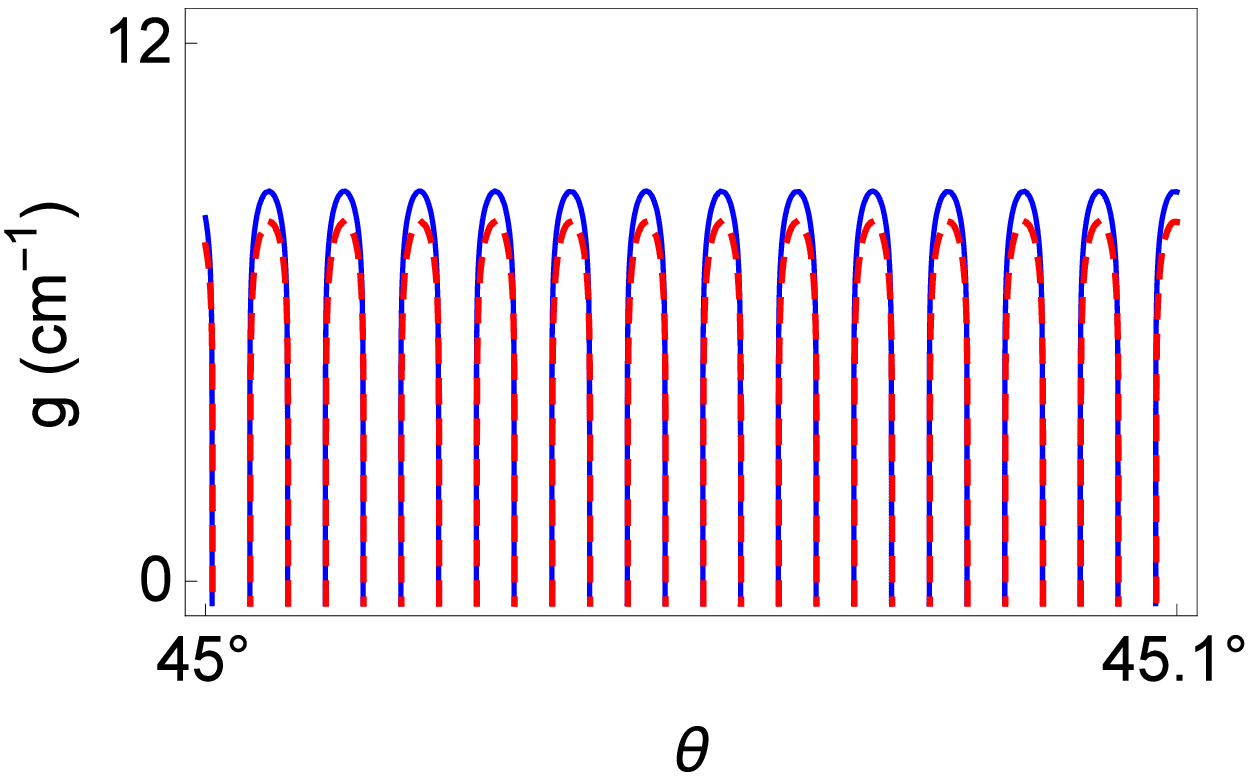}~~~
    \includegraphics[scale=.60]{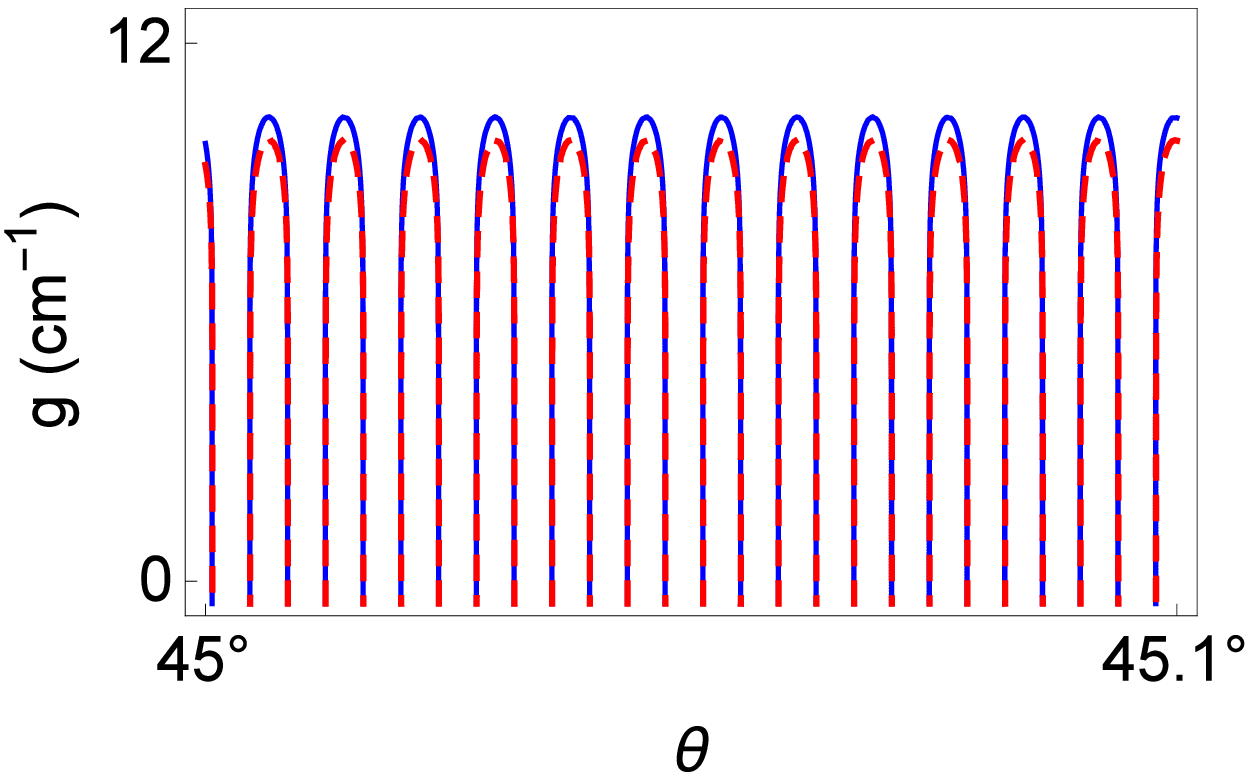}\\
    \includegraphics[scale=.60]{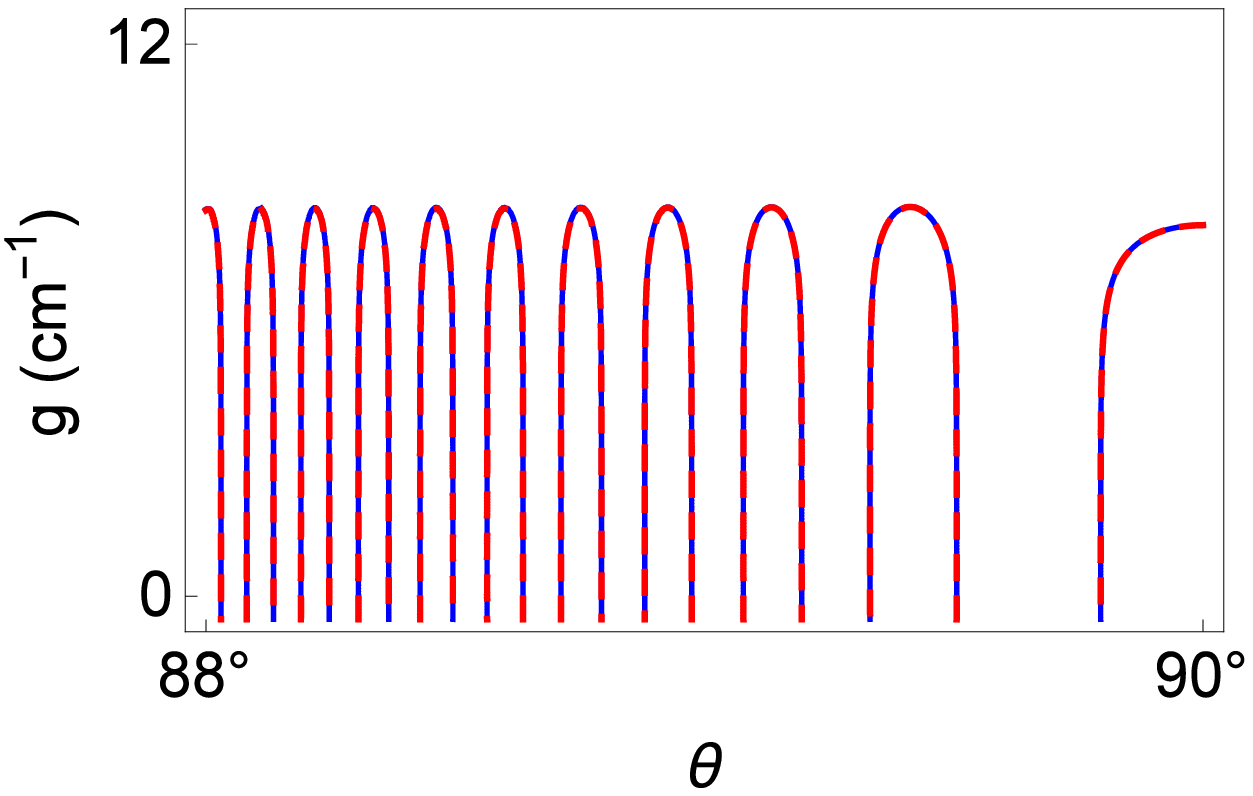}~~~
    \includegraphics[scale=.60]{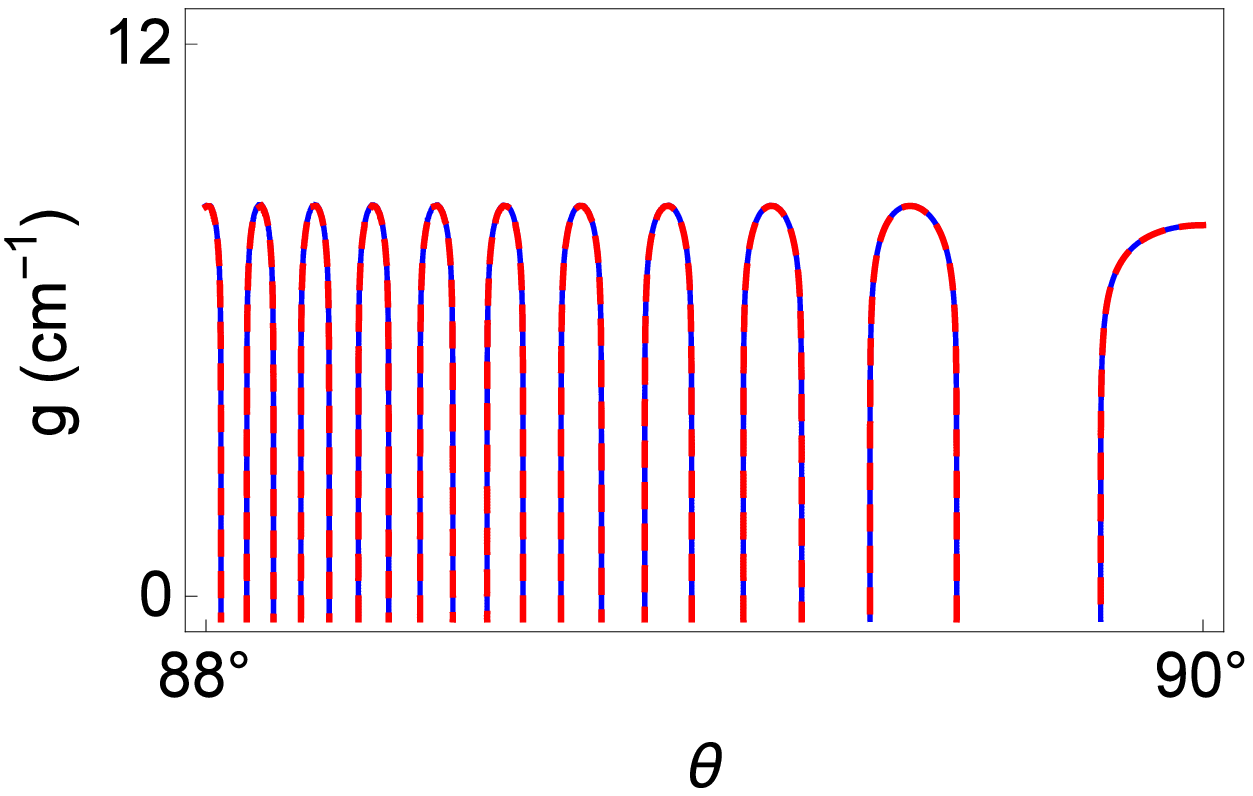}
    \caption{(Color online) Plots of gain coefficients as a function of incidence angle $\theta$. Solid blue curves reflect the situation in the absence of graphene while dashed red curves yield the states with graphene. As the incidence angle rises, gain values corresponding to the left and right reflectionlessness approach to each other, and effect of graphene vanishes. Graphene is highly effective at small angles.}
    \label{fig6}
    \end{center}
    \end{figure}

Finally, Fig.~\ref{fig6} explicitly reveals the graphical demonstrations of how gain coefficient $g$ depends upon the incidence angle $\theta$. Notice that not all incidence angles provide a unidirectional reflectionlessness, but it occurs at periodically decreasing angle ranges. These plots exhibit that existence of graphene reduces gain value while it shifts incidence angle slightly at fixed wavelength $\lambda$. Again we see that gain value for the right reflectionlessness is higher compared to the left one. More importantly, for large angles the effect of graphene almost disappears and the same gain values leads to a bidirectional configurations.

Next, we keep track of the similar approach of the previous argument for the perturbative expression for the unidirectional invisibility and thus employ Eqs.~20 - 23 in (19) in the main text and in turn decompose into the real and imaginary parts. This gives rise to the following set of equations
    \begin{align}
    \cX_{+}^{(\ell)} e^{\ell \tilde{g}L} - \cX_{-}^{(\ell)} e^{-\ell \tilde{g}L} =& \frac{2\tilde{\eta}\sin\left(2\ell k_zL\right)}{\tilde{\kappa}}\left[\alpha_{\ell}\cos\left(2\ell k_zL\right)-\beta_{\ell}\sin\left(2\ell k_zL\right)\right]\notag\\ &+ 2 \left[\alpha_{\ell}\cos\left(4\ell k_zL\right)-\beta_{\ell}\sin\left(4\ell k_zL\right)\right], \label{realeqn}\\
    \cW_{+}^{(\ell)} e^{\ell \tilde{g}L} + \cW_{-}^{(\ell)} e^{-\ell \tilde{g}L} =& \frac{2\tilde{\eta} \cos\left(2\ell k_zL\right)}{\tilde{\kappa}}\left[\beta_{\ell}\sin\left(2\ell k_zL\right)-\alpha_{\ell}\cos\left(2\ell k_zL\right)\right]\notag\\ &- 2 \left[\alpha_{\ell}\sin\left(4\ell k_zL\right)+\beta_{\ell}\cos\left(4\ell k_zL\right)\right], \label{imageqn}
    \end{align}
where we use the following identifications for convenience
   \begin{align}
   \alpha_{\ell} &:= -2\,\textrm{Re}[\fu_{\ell}]\textrm{Im}[\fu_{\ell}],~~~~~~~~~~ \beta_{\ell} := 1 - \textrm{Re}^2[\fu_{\ell}] + \textrm{Im}^2[\fu_{\ell}], \notag\\
   \cX_{\pm}^{(\ell)} &:= 2\,\textrm{Im}[\fu_{\ell}] \left(1\pm \textrm{Re}[\fu_{\ell}]\right)\cos(2\ell k_zL\tilde{\eta}) + \left[\left(1 \pm \textrm{Re}[\fu_{\ell}]\right)^2 - \left(\textrm{Im}[\fu_{\ell}]\right)^2\right]\sin(2\ell k_zL\tilde{\eta}), \notag\\
   \cW_{\pm}^{(\ell)}&:= 2\,\textrm{Im}[\fu_{\ell}] \left(1\pm \textrm{Re}[\fu_{\ell}]\right)\sin(2\ell k_zL\tilde{\eta}) - \left[\left(1 \pm \textrm{Re}[\fu_{\ell}]\right)^2 - \left(\textrm{Im}[\fu_{\ell}]\right)^2\right]\cos(2\ell k_zL\tilde{\eta}), \notag
   \end{align}
and again the index $\ell = + / -$ represents the left/right invisible configurations. Eqs. \ref{realeqn} and \ref{imageqn} are transcendental equations that lead to ultimate solutions among the parameters of the optical system. Notice that the effect of graphene is revealed in the presence of $\fu_{\ell}$. When the graphene layers are removed so that $\fu_{\ell} = \tilde{\fn}_{\ell}^{-1}$, Eq.~\ref{realeqn} vanishes and parameters of the system are restricted to Eq.~\ref{imageqn} which yields Eq.~27 in \cite{pra-2017a}.

\begin{figure}
    \begin{center}
    \includegraphics[scale=.70]{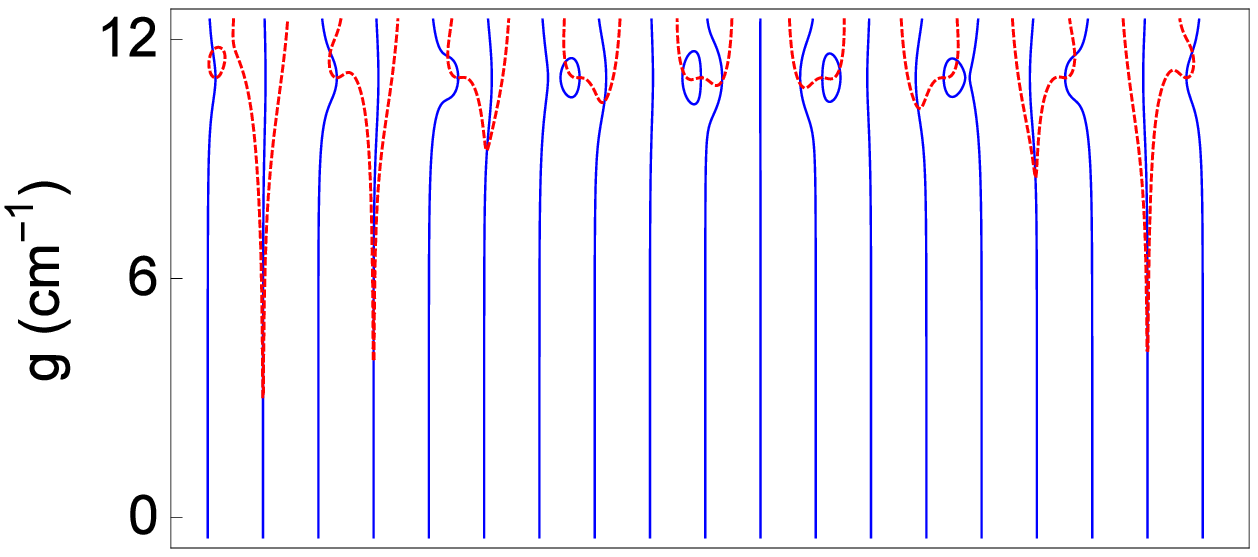}\\
    ~~~\includegraphics[scale=.745]{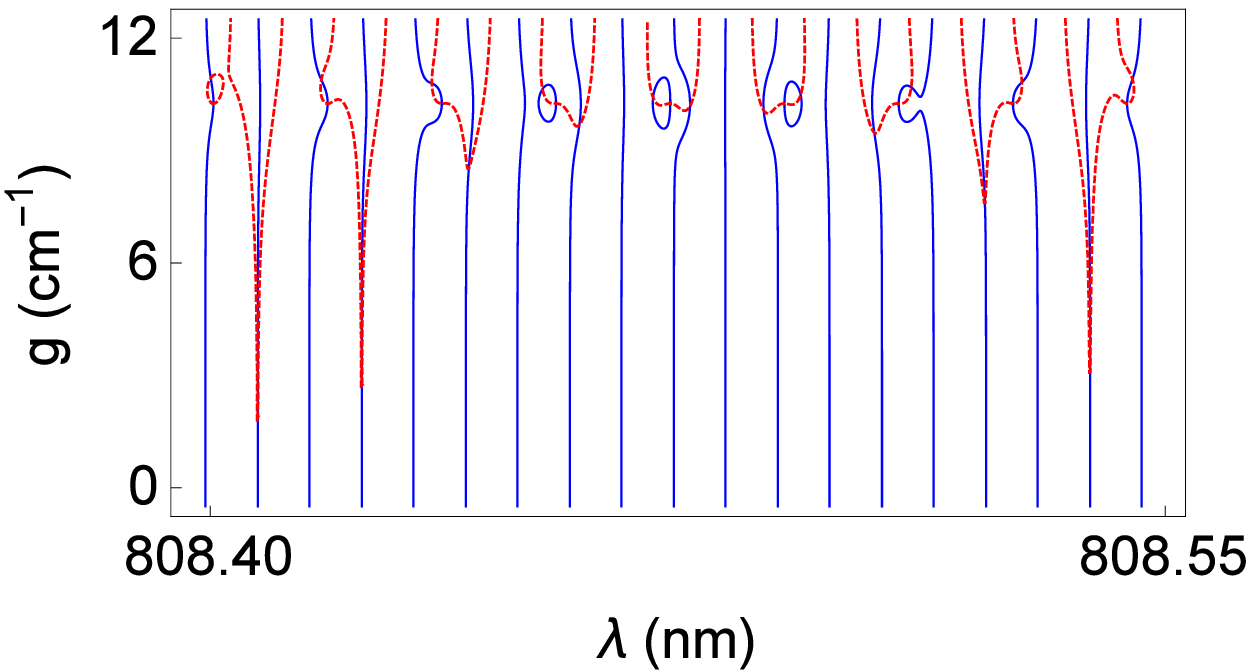}
    \caption{(Color online) Plots displaying the effect of graphene over unidirectional invisibility. Top row describes grapheneless structure while bottom row belongs to $\cP\cT$-symmetric structure with graphene as given by the properties in (25) and (26) in the main text. Solid blue curves represent Eq.~\ref{realeqn} and dashed red curves Eq.~\ref{imageqn}. Invisible patterns are obtained along intersection points.}
    \label{fig7}
    \end{center}
    \end{figure}

Let us demonstrate the concrete physical realizations of Eqs.~\ref{realeqn} and \ref{imageqn} for the invisibility phenomenon by using the sample optical configuration in (25) and (26) in the main text. In Fig.~\ref{fig7}, we illustrate the effect of graphene by means of parameter values in (25) and (26). Upper figure stands for the grapheneless pure $\cP\cT$-symmetric optical structure for the right invisibility, which is attained by means of (25) and (26) in the main text. Solid blue curves represent the real equation given in (\ref{realeqn}) and dashed red curves denote (\ref{imageqn}). Intersection of these curves gives rise to the invisible configurations that we seek out. In fact, this case has been analyzed in \cite{pra-2017a} in detail. The lower figure, on the other hand, brings out the influence of having graphene over $\cP\cT$-symmetric structure. As we pointed out in reflectionless case, the presence of graphene yields the invisible gain value to lower, and the wavelength interval to right-shift slightly. This means that we can obtain invisible configurations using smaller gain values compared to the grapheneless structures by possessing large chemical potentials and small temperatures. Thus, we can observe invisibility at any wavelength range as we desire, just as pointed out in Figs.~\ref{fig3} and \ref{fig4}.

\begin{figure}
    \begin{center}
    \includegraphics[scale=.70]{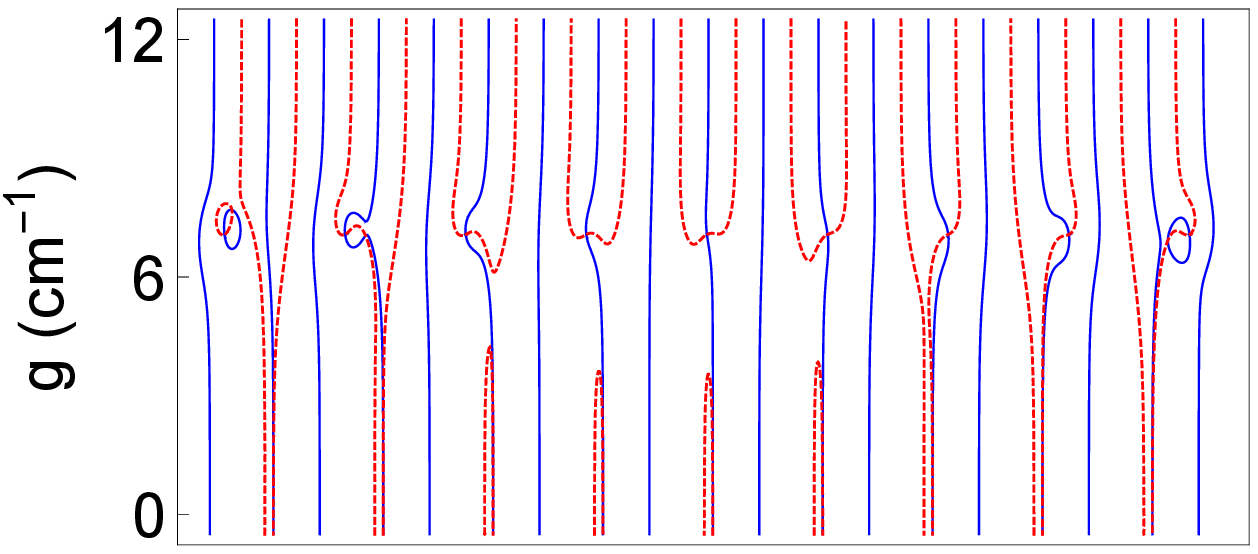}\\
    ~~~\includegraphics[scale=.745]{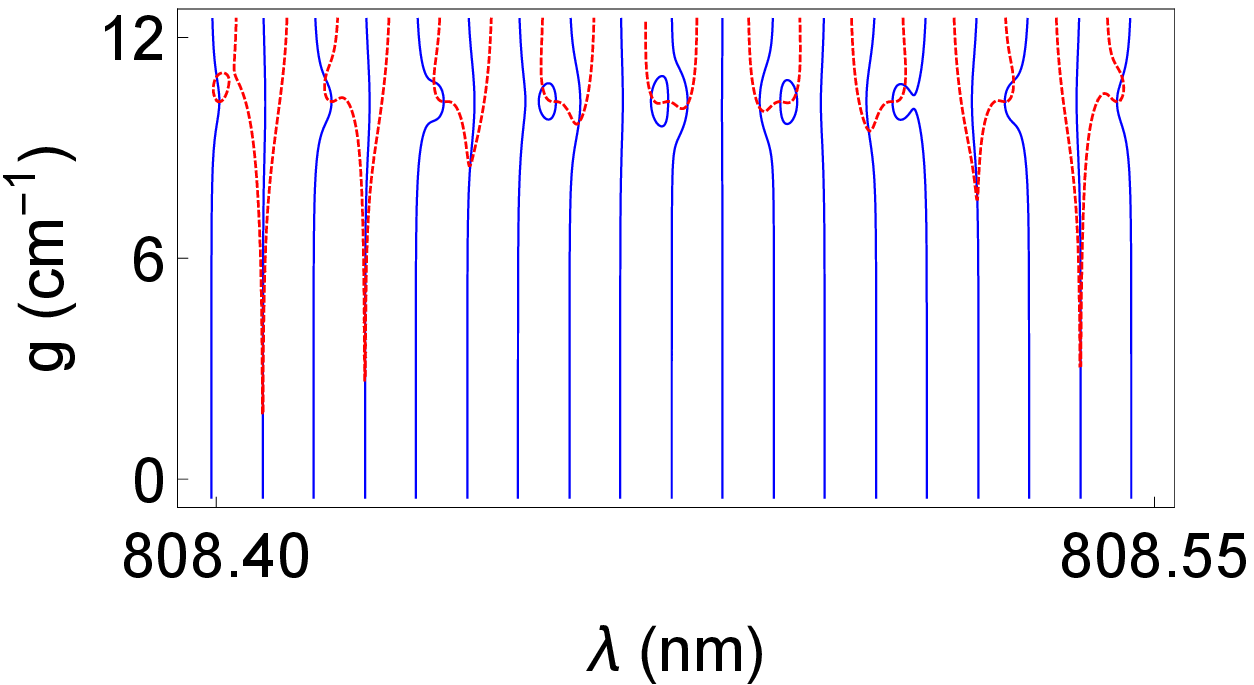}
    \caption{(Color online) Comparison of left (top figure) and right (bottom figure) invisibility over gain coefficient $g$ and wavelength $\lambda$ graph in case of graphene. Invisible patterns are represented by intersection points.}
    \label{fig8}
    \end{center}
    \end{figure}

In Fig.~\ref{fig8}, we compare the effect of graphene over left and right invisible configurations using parameter specifications given in (25) and (26). Solid blue and dashed red curves correspond to (\ref{realeqn}) and (\ref{imageqn}) respectively for the case of $\ell = +$. Thus, invisibility shows up to exist at intersection points. We observe that one requires higher gain amounts for the right invisibility compared to the left one, and at certain wavelengths left and right invisibility gain amounts overlap, which produce bidirectional invisibility. Therefore, one can deduce that if higher chemical potential and lower temperature is opted for graphene in a relatively smaller size of slab layers respecting $\cP\cT$-symmetry, then rather small amount of gain is enough to observe uni- or bidirectional invisibility at any wavelength range in demand.

In Figs.~\ref{figpi1}-\ref{figpi5}, we use Nd:YAG crystals with specifications given in (25) and graphene layers in (26) in the main text, and represent the quantities $\left|R^{l}\right|^2$, $\left|R^{r}\right|^2$ and $\left|T-1\right|^2$ by thin solid red, thick dashed blue and thin dashed green curves respectively. Nonintersecting dashed blue and solid red curves denote the unidirectional reflectionless configurations, whose intersections thus amount to bidirectional reflectionless configurations. Intersection of either one of these curves with the thin dashed green curves simply means that a unidirectional invisibility occurs at the corresponding overlap regions. In the following Figs.~\ref{figpi1}-\ref{figpi5}, We sort out various situations regarding associated parameters which are distinguishable in aspects of reflectionlessness and invisibility phenomena.

    \begin{figure}
    \begin{center}
    \includegraphics[scale=.6]{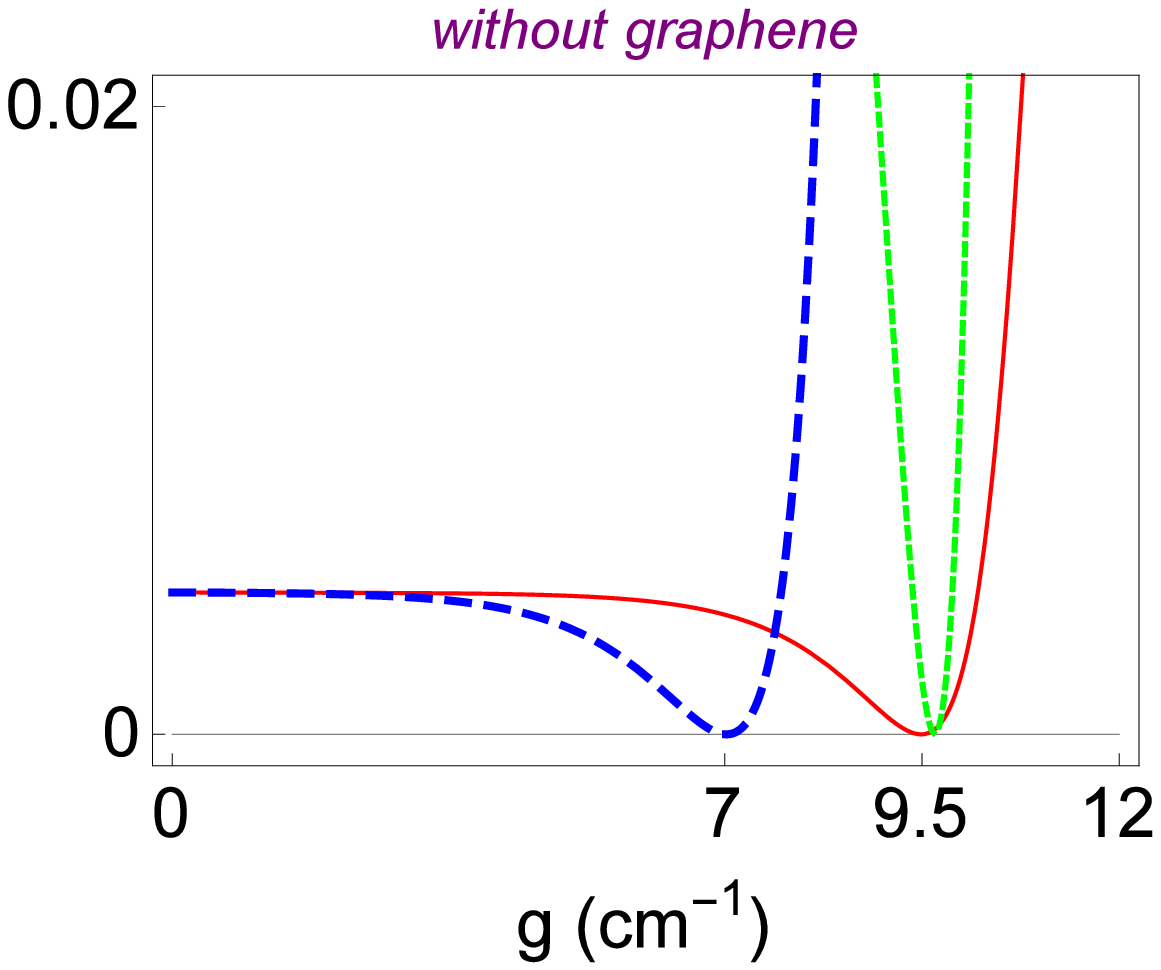}~~~
    \includegraphics[scale=.6]{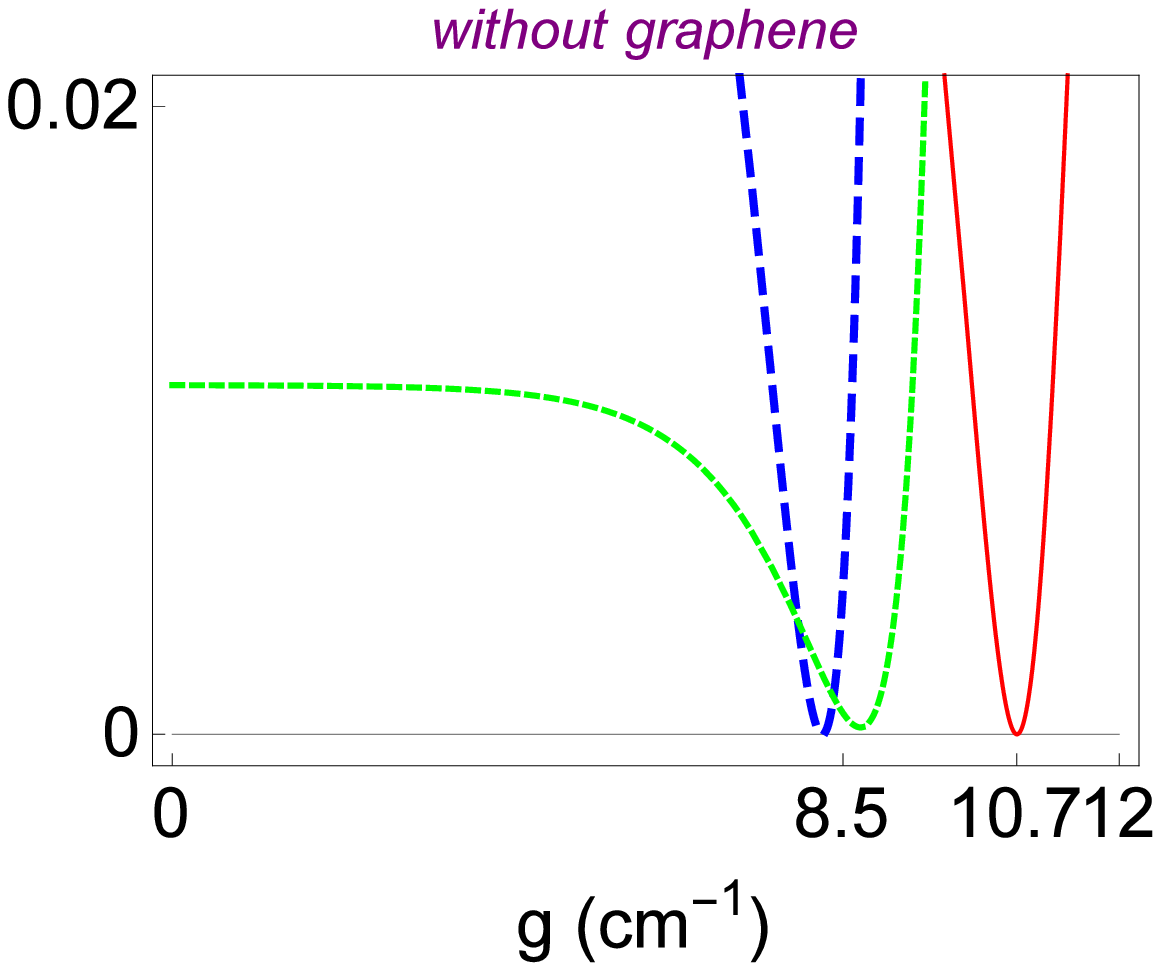}\\
    \includegraphics[scale=.6]{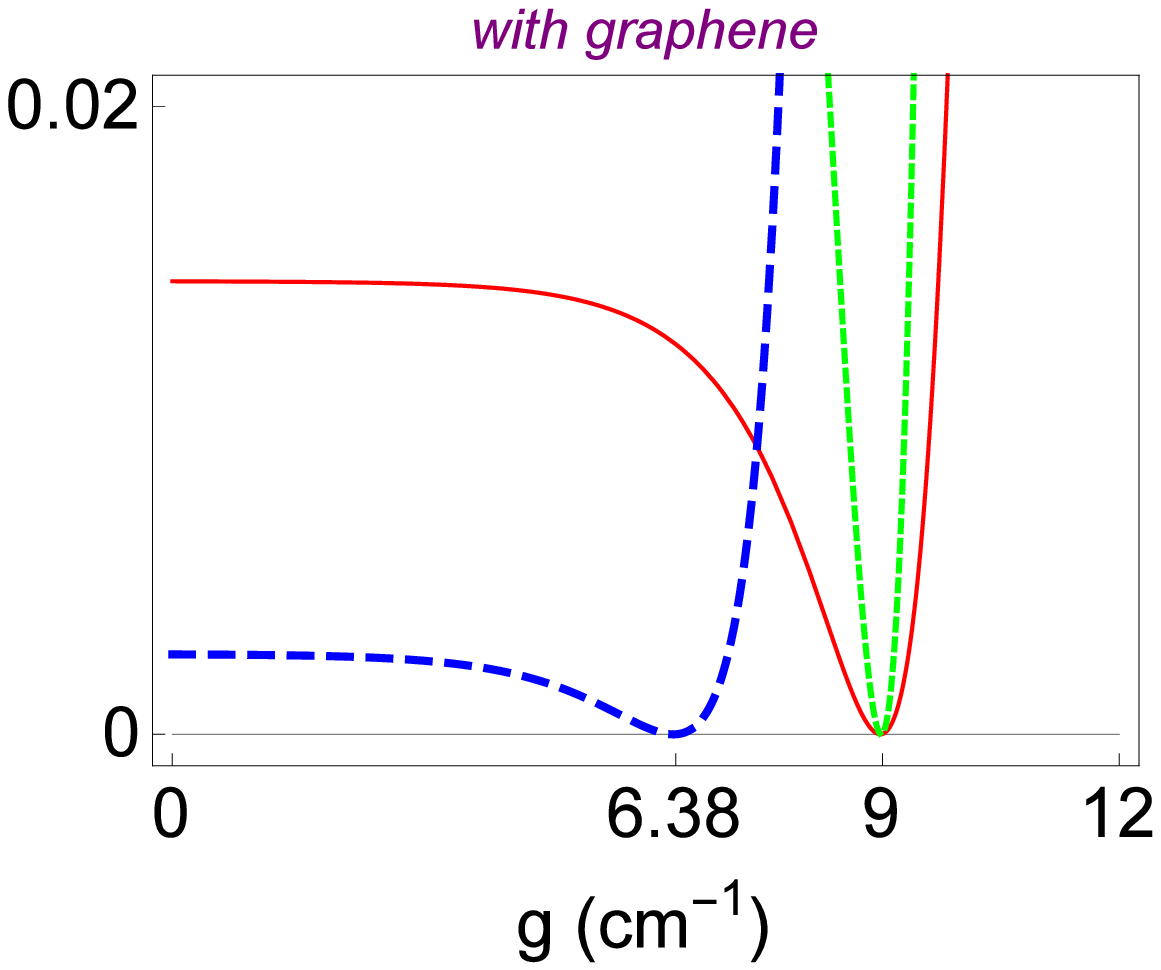}~~~
    \includegraphics[scale=.6]{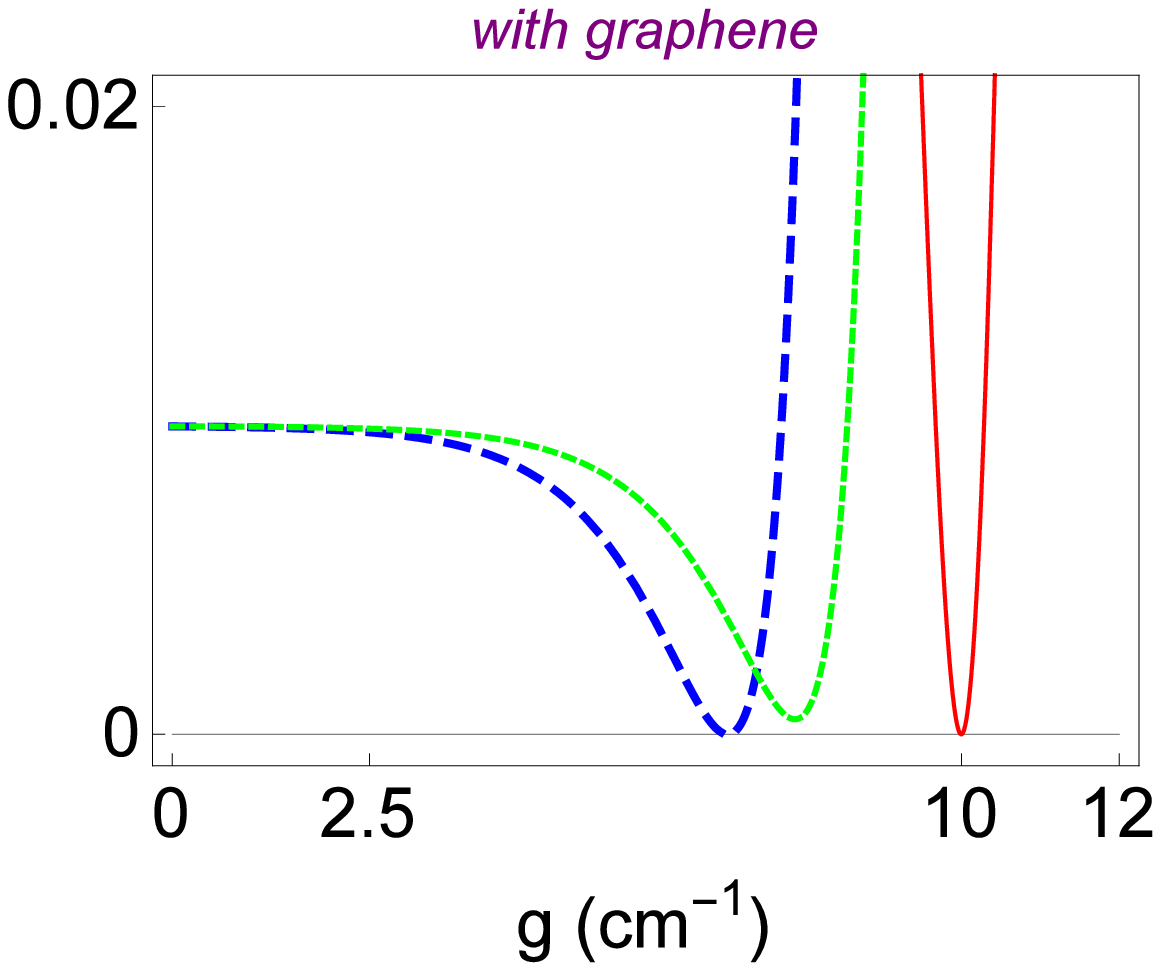}
    \caption{(Color online) Plots of $\left|\textrm{R}^{l}\right|^2$ (thick dashed blue curve), $\left|\textrm{R}^{r}\right|^2$ (thin red curve) and $\left|\textrm{T}-1\right|^2$ (thin dashed green curve) displayed on the vertical axis as a function of gain amount $g$ with and without graphenes. It is clearly seen that required gain values for uni- or bidirectional reflectionlessness and invisibility are reduced significantly.}
    \label{figpi1}
    \end{center}
    \end{figure}
Fig.~\ref{figpi1} clarifies the role of graphene in minimizing the gain value in the $\cP\cT$-symmetric slab system. We situate the slab parameters as in (25) with different choices of incidence angles, and (26) as graphene specifications. Figures on the left/right column have incidence angle of $\theta =1.313^{\circ}$/$\theta =1.305^{\circ}$ and wavelength $\lambda = 808~\textrm{nm}$ with and without graphenes, and represent the prescribed phenomena within $2\%$ of illusion. Upper left figure is bidirectionally reflectionless up to the gain value $g\approx 3~\textrm{cm}^{-1}$, right reflectionless between gain values $g\approx 3~\textrm{cm}^{-1}$ and $g\approx 10.5~\textrm{cm}^{-1}$, left reflectionless between $g\approx 3~\textrm{cm}^{-1}$ and $g\approx 8~\textrm{cm}^{-1}$ and right invisible around the gain value of $g = 9.65~\textrm{cm}^{-1}$. On the other hand, presence of graphene in the lower left figure removes long range bidirectional reflectionlessness and spots it to just one point at $g = 7.4~\textrm{cm}^{-1}$, the range of left and right reflectionlessness increases, and right invisibility shifts to a smaller gain value at $g= 9~\textrm{cm}^{-1}$. Likewise, upper and lower right figures unveil that whereas there is no left invisibility except for two points in absence of graphene, it occurs till the gain value $g = 2.5~\textrm{cm}^{-1}$ when the graphene sheets are located. Similarly, left and right reflectionless gain values get smaller with the presence of graphene, see Fig.~\ref{figpi1}.

    \begin{figure}
    \begin{center}
    ~~~~~~\includegraphics[scale=.56]{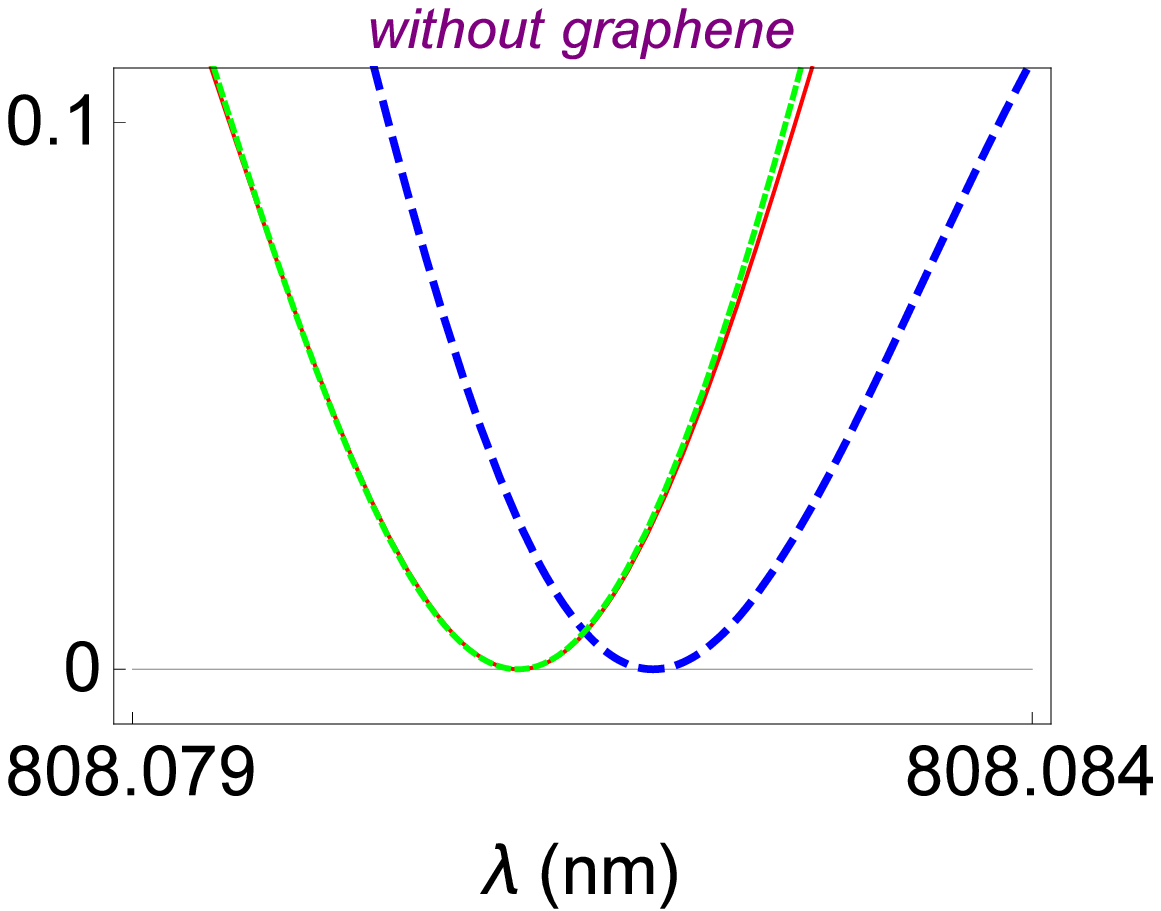}~~~~~~~
    \includegraphics[scale=.6]{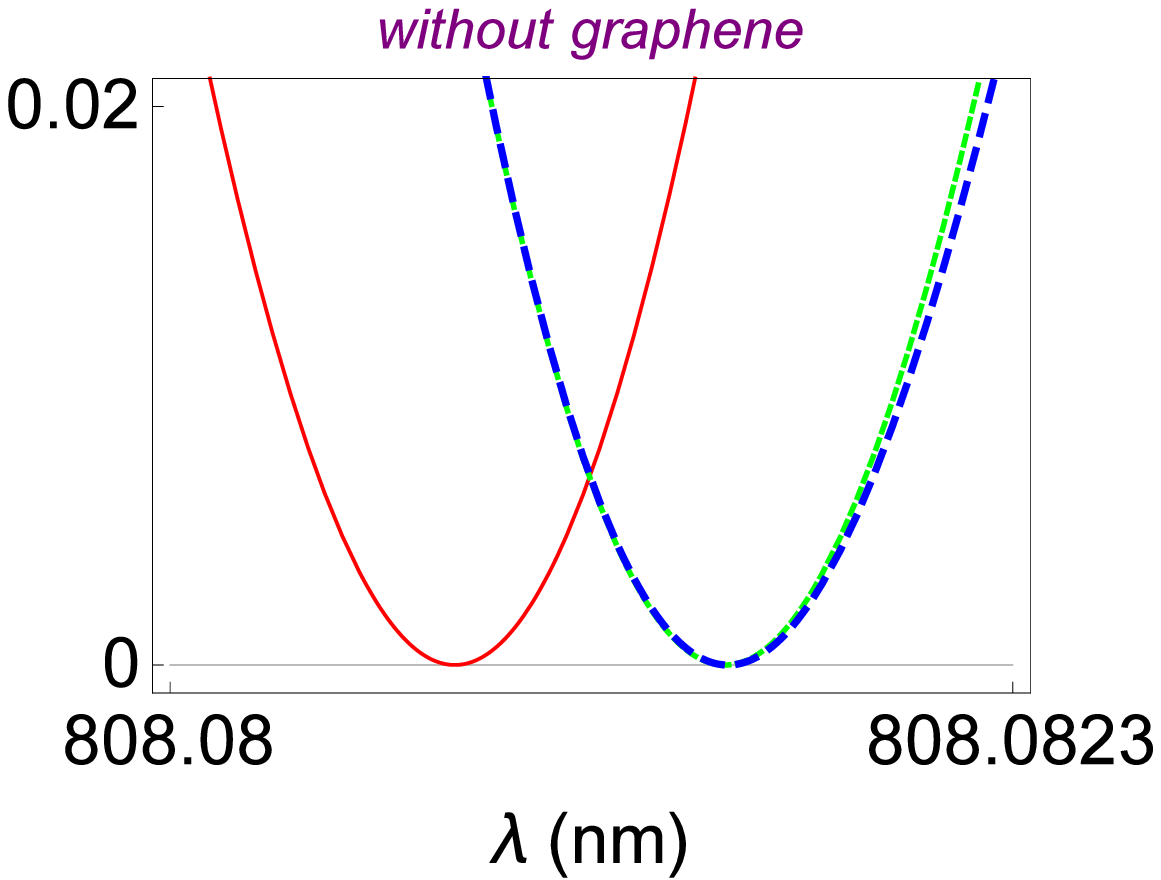}\\
    \includegraphics[scale=.65]{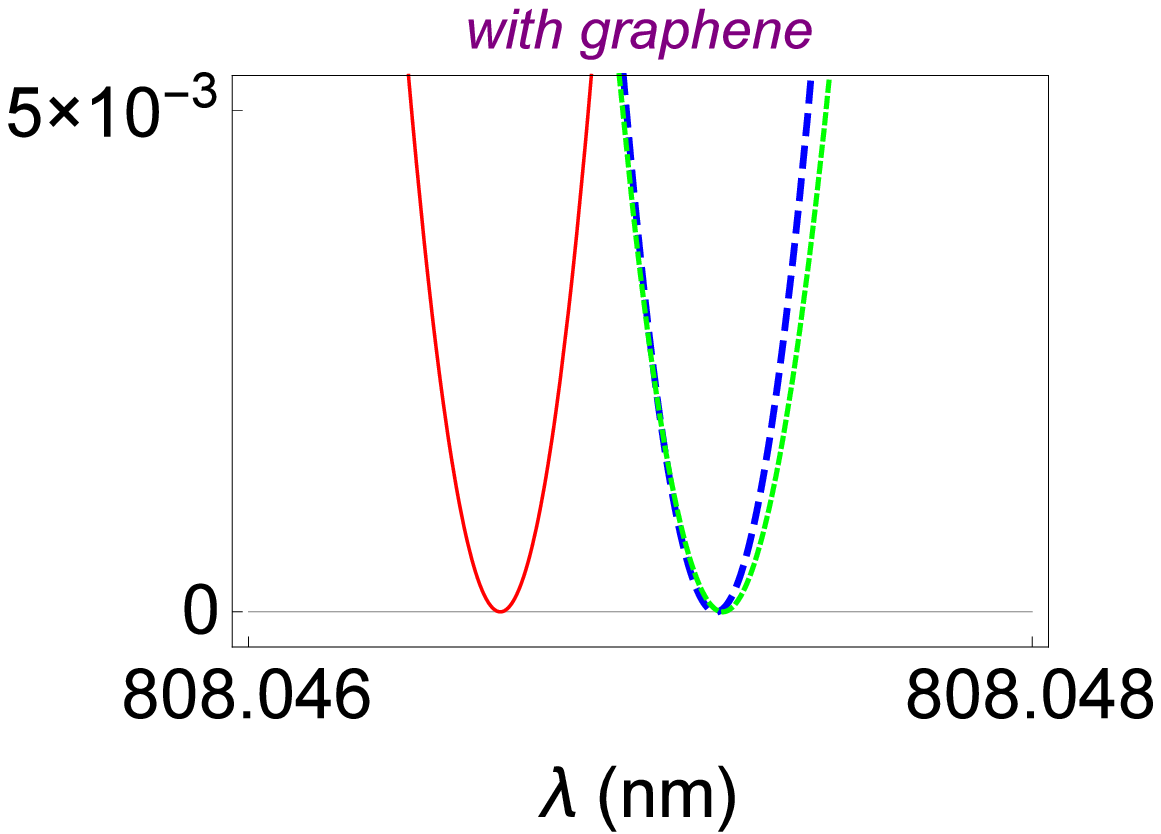}~
    \includegraphics[scale=.65]{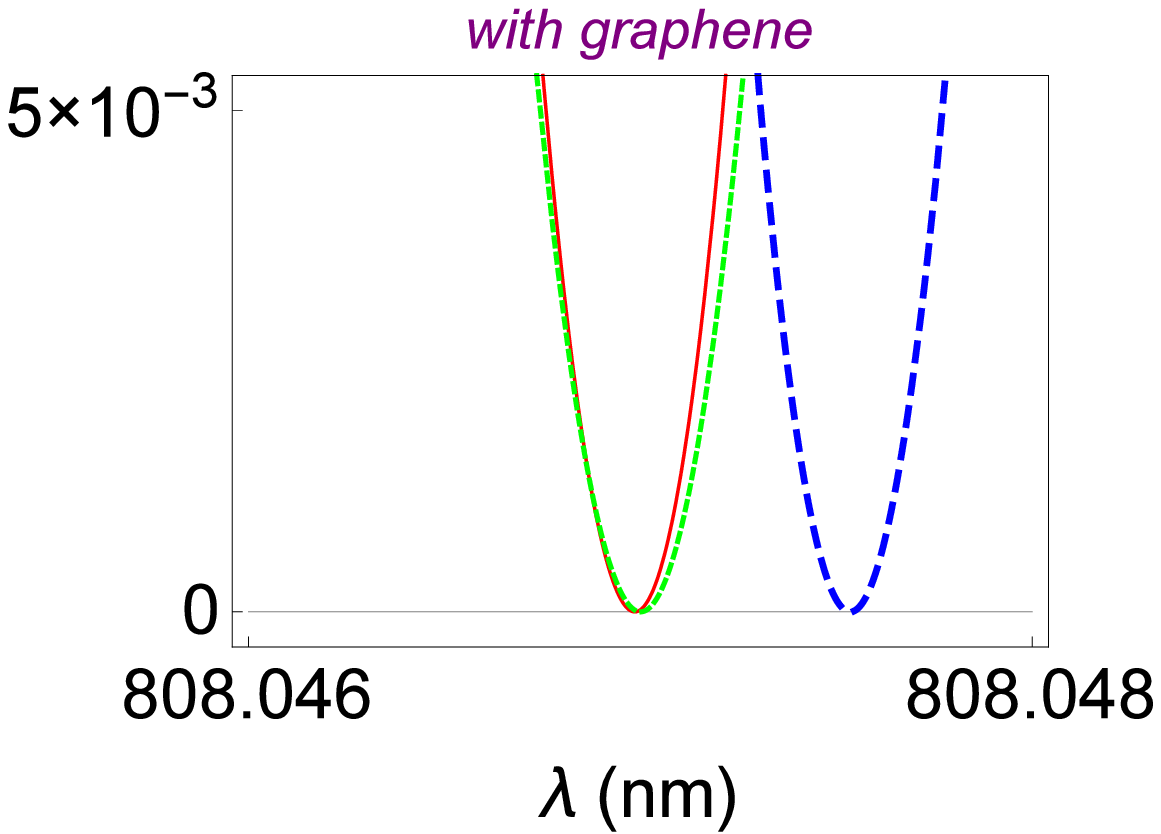}
    \caption{(Color online) Plots of $\left|\textrm{R}^{l}\right|^2$ (thick dashed blue curve), $\left|\textrm{R}^{r}\right|^2$ (thin red curve) and $\left|\textrm{T}-1\right|^2$ (thin dashed green curve) displayed on the vertical axis as a function of wavelength $\lambda$ with/ and without graphenes. It is revealed that left invisibility turns into the right one, and vice versa.}
    \label{figpi2}
    \end{center}
    \end{figure}
Fig.~\ref{figpi2} illustrates the influence of graphene on the wavelength dependence of quantities $\left|\textrm{R}^{l}\right|^2$, $\left|\textrm{R}^{r}\right|^2$ and $\left|\textrm{T}-1\right|^2$. We employ the gain value of $g = 8~\textrm{cm}^{-1}$. In the absence of graphene in upper left figure, one observes the right invisibility and left reflectionlessness squeezed in the wavelength interval of $(808.079~\textrm{nm}, 808.084~\textrm{nm})$. With the insertion of graphene displayed in lower left figure, the same patterns turn into the left invisibility and right reflectionlessness in the new shifted wavelength interval of $(808.046~\textrm{nm}, 808.048~\textrm{nm})$. We also note that incidence angle shifts to the right from $\theta = 1.312^{\circ}$ to $\theta = 1.327^{\circ}$.  Similarly, right reflectionlessness and left invisibility in lack of graphene as shown in upper right figure are restricted to the wavelength interval $(808.08~\textrm{nm}, 808.0823~\textrm{nm})$ at incidence angle $\theta = 1.315^{\circ}$. The use of graphene yields right invisibility and left reflectionlessness at incidence angle $\theta = 1.324^{\circ}$. Thus we conclude that graphene results in the left-shifting of wavelength range of reflectionlessness and invisibility, together with right shifting the angle of incidence.

    \begin{figure}
    \begin{center}
    \includegraphics[scale=.4]{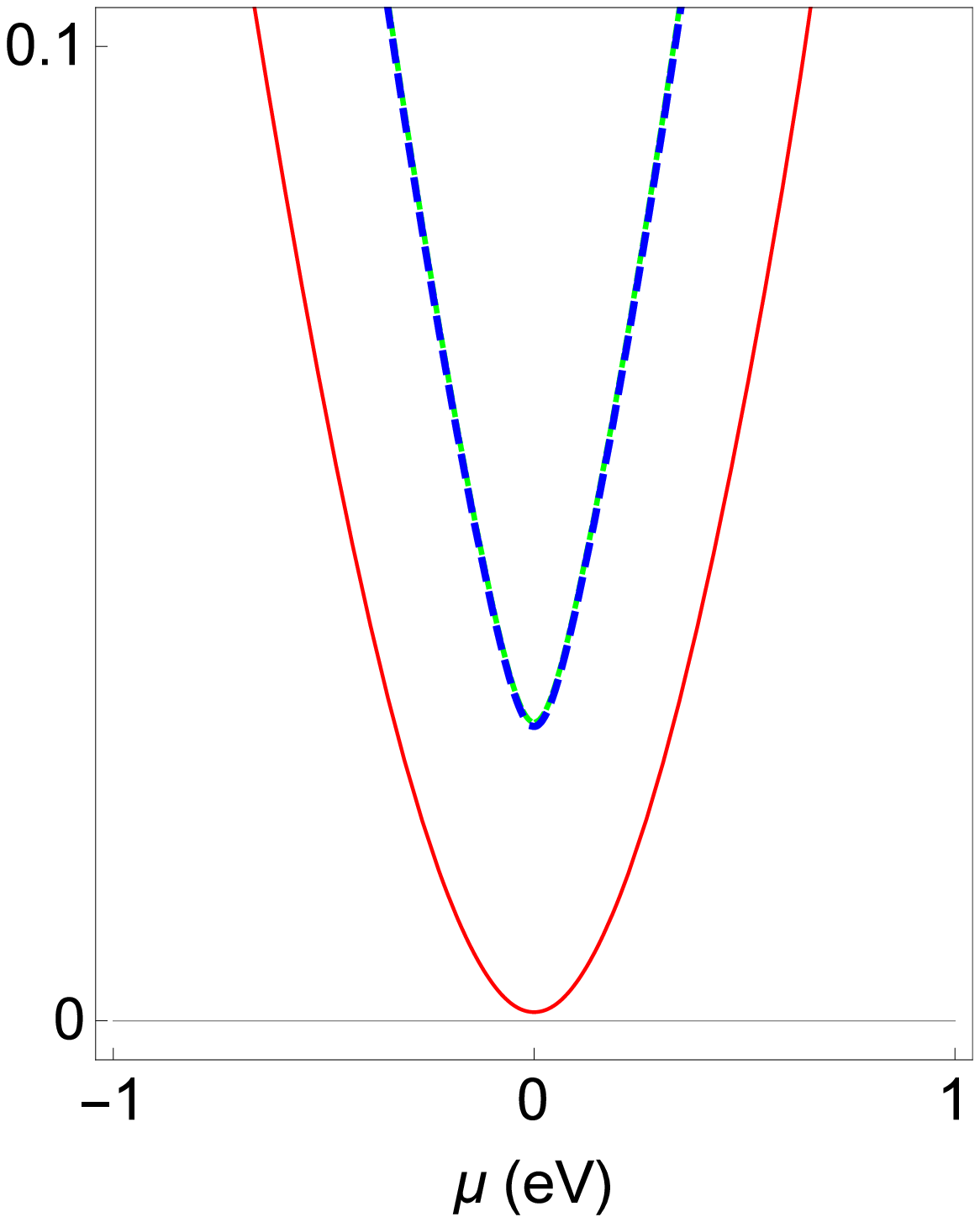}~
    \includegraphics[scale=.4]{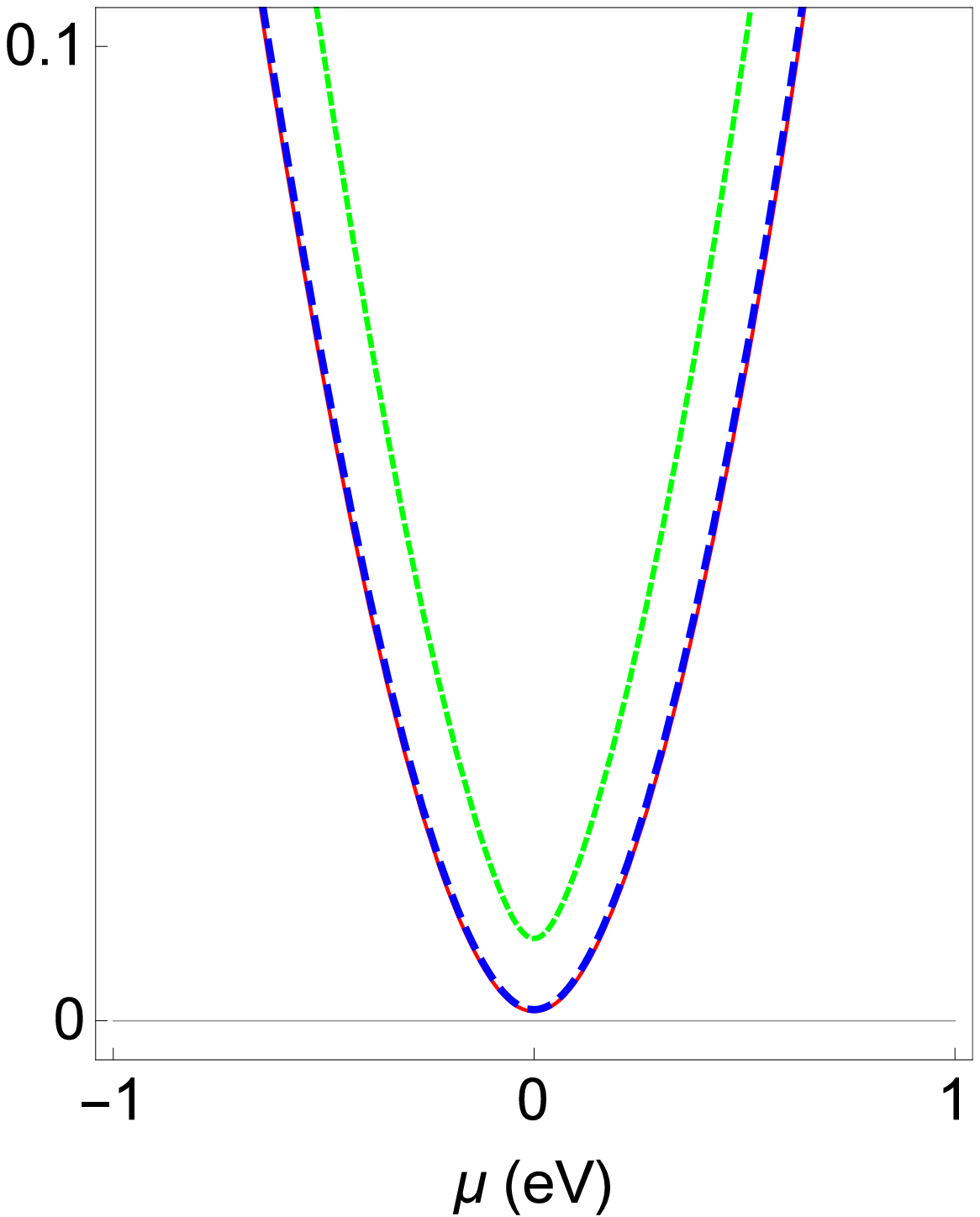}~
    \includegraphics[scale=.4]{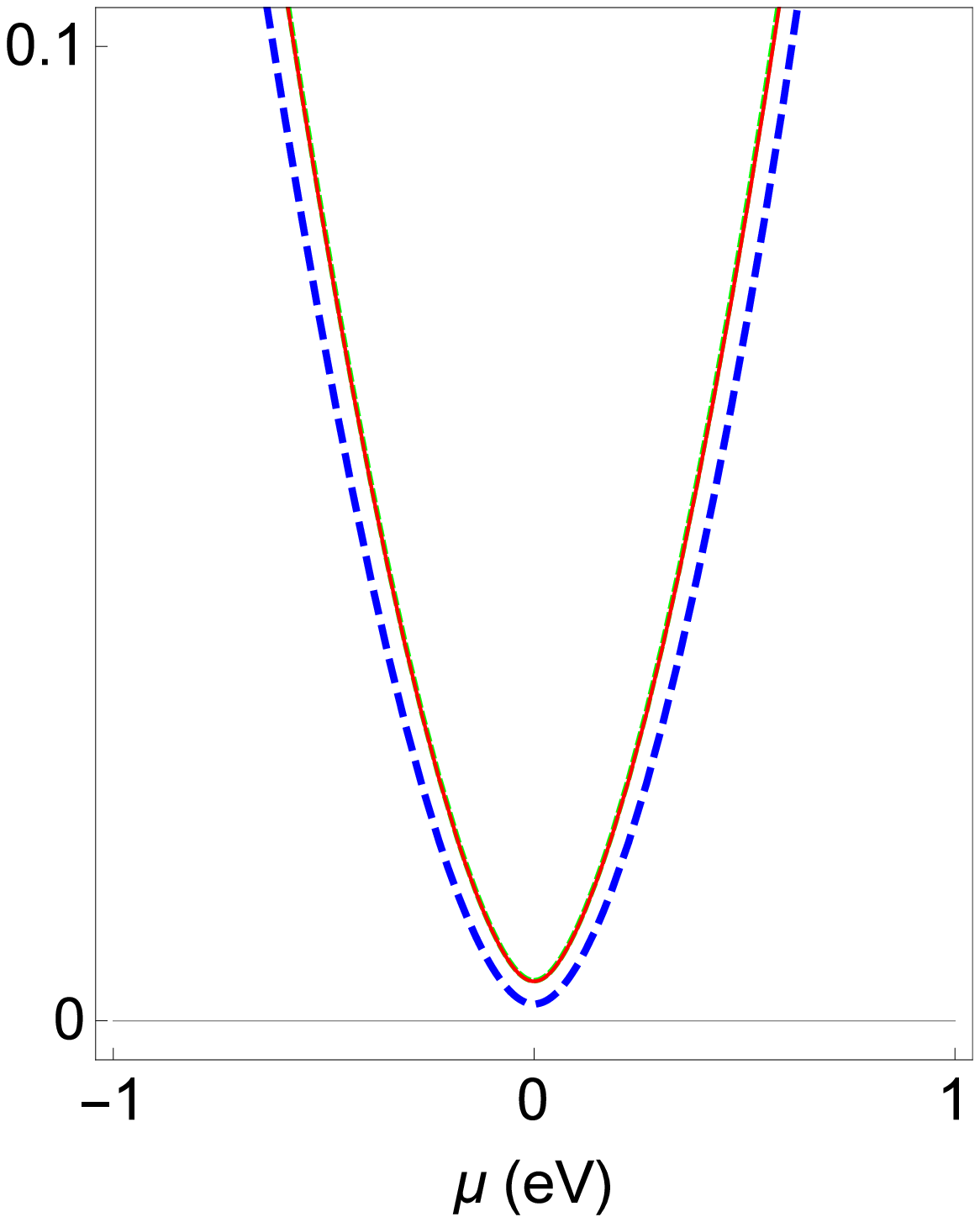}
    \caption{(Color online) Plots of $\left|\textrm{R}^{l}\right|^2$ (thick dashed blue curve), $\left|\textrm{R}^{r}\right|^2$ (thin red curve) and $\left|\textrm{T}-1\right|^2$ (thin dashed green curve) displayed on the vertical axis as a function of chemical potential $\mu$. Level of Unidirectional reflectionlessness and invisibility increases with the decrease of $\mu$ around $\mu=0~\textrm{eV}$.}
    \label{figpi3}
    \end{center}
    \end{figure}
Fig.~\ref{figpi3} clearly demonstrates the effect of chemical potential variation of graphene sheets on the quantities $\left|\textrm{R}^{l}\right|^2$, $\left|\textrm{R}^{r}\right|^2$ and $\left|\textrm{T}-1\right|^2$ for the Nd:YAG crystals. We fix gain value to $g= 5~\textrm{cm}^{-1}$ and wavelength to $\lambda= 808.081~\textrm{nm}$. We immediately observe that all unidirectional reflectionless and invisible patterns take place at small chemical potentials which are typically less than $\left|\mu\right|=1~\textrm{eV}$ within $1\%$ of flexibility. At incidence angle of $\theta = 1.315^{\circ}$ in the left figure, one gets right reflectionless and left invisible configurations. A slight change of incidence angle to $\theta = 1.314^{\circ}$ in the middle figure yields bidirectional configuration and finally incidence angle of $\theta = 1.313^{\circ}$ in the right figure gives left reflectionless and right invisible configurations. As a result, reducing chemical potential very close to zero gives rise to quite reliable patterns.

    \begin{figure}
    \begin{center}
    \includegraphics[scale=.5]{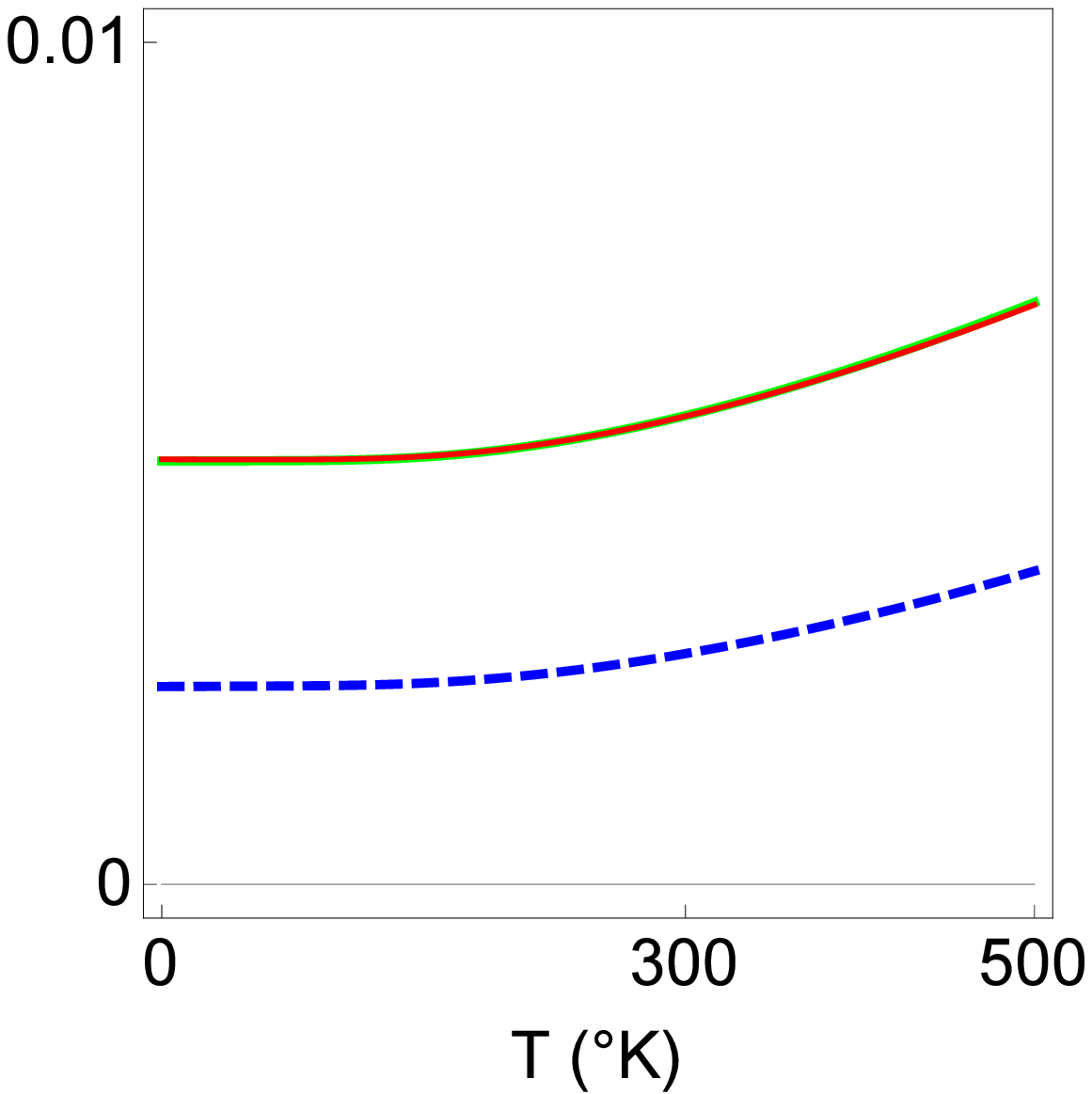}~~
    \includegraphics[scale=.5]{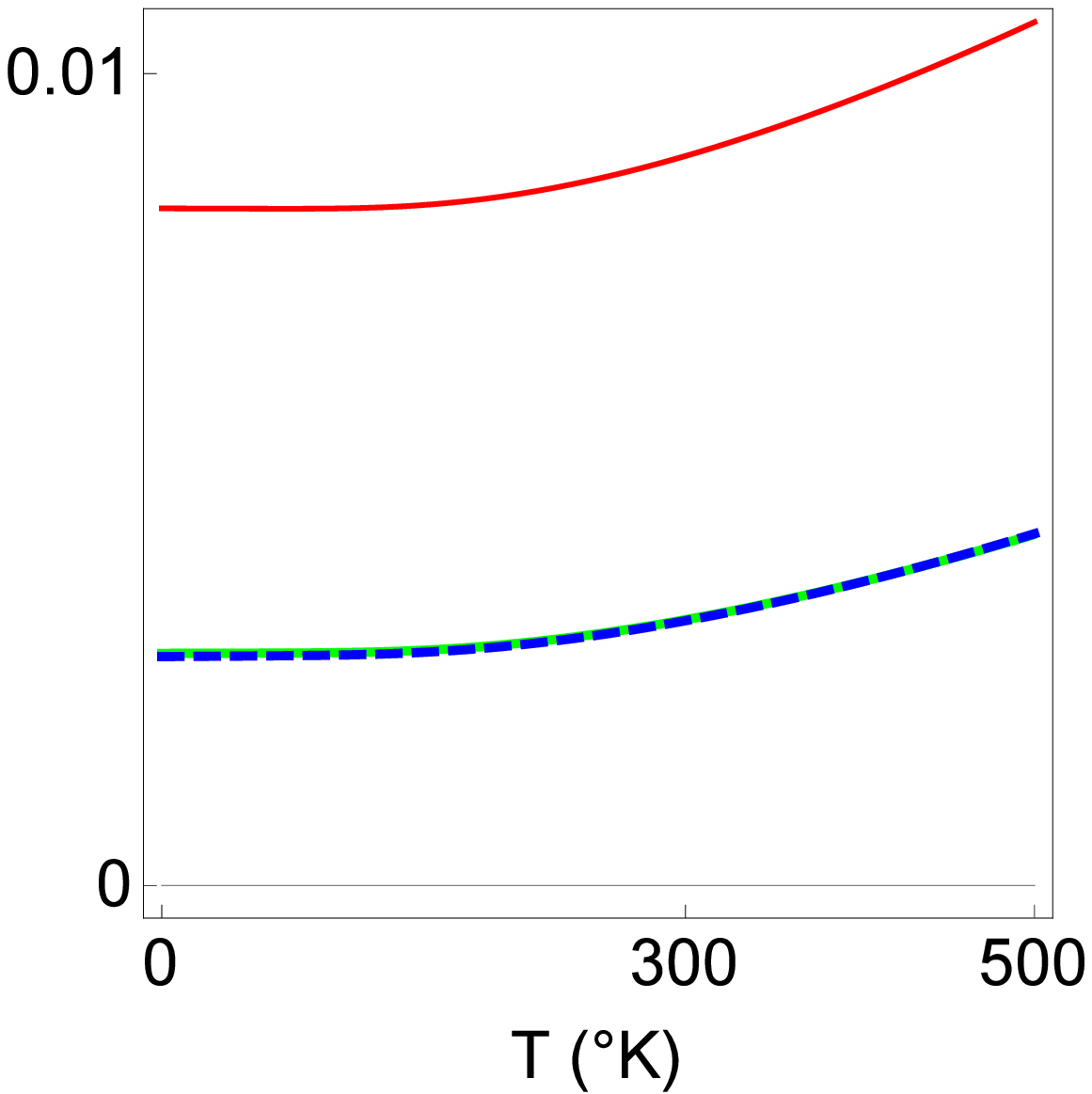}
    \caption{(Color online) Figure showing the temperature variations of $\left|\textrm{R}^{l}\right|^2$ (thick dashed blue curve), $\left|\textrm{R}^{r}\right|^2$ (thin red curve) and $\left|\textrm{T}-1\right|^2$ (thin dashed green curve). Temperature change does not induce losing the type of reflectionlessness and invisibility, but just yields a better precision of them.}
    \label{figpi4}
    \end{center}
    \end{figure}
Fig.~\ref{figpi4} displays the impact of temperature variations of graphene sheets over the reflectionless and invisible structures. We again employ the gain value of $g= 5~\textrm{cm}^{-1}$ and wavelength $\lambda= 808.081~\textrm{nm}$. We reveal that almost all accessible temperature values produce left and right reflectionless and invisible patterns, but smallest possible temperatures give rise to the best precise configurations.

    \begin{figure}
    \begin{center}
    \includegraphics[scale=.6]{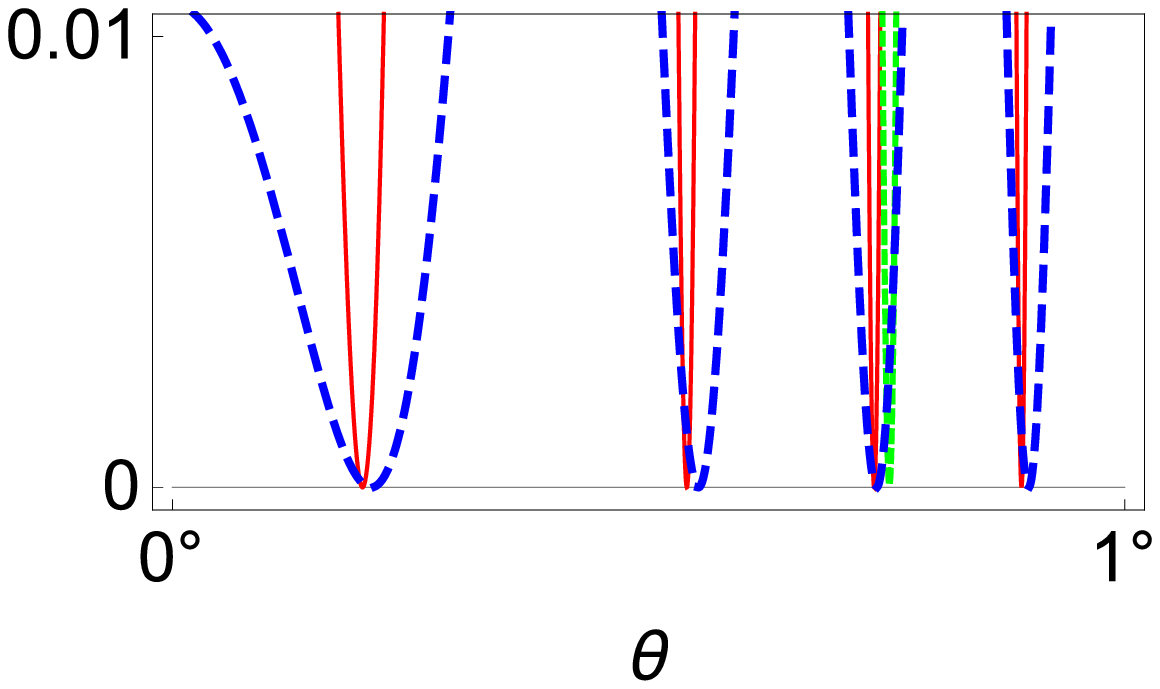}
    \includegraphics[scale=.63]{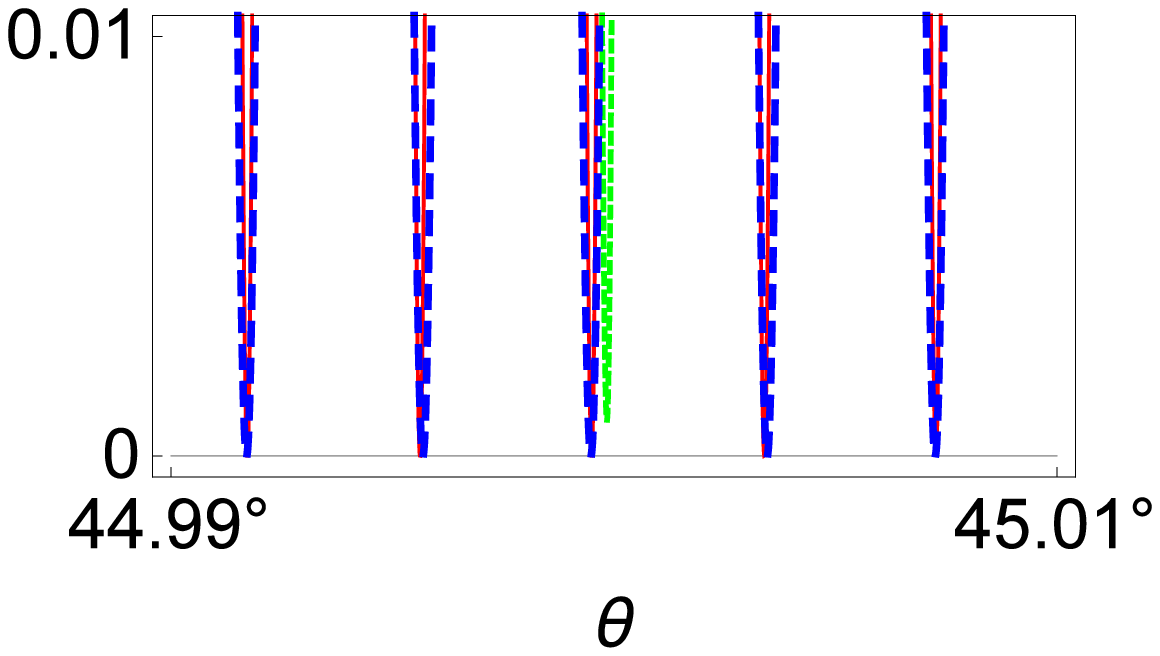}
    \includegraphics[scale=.6]{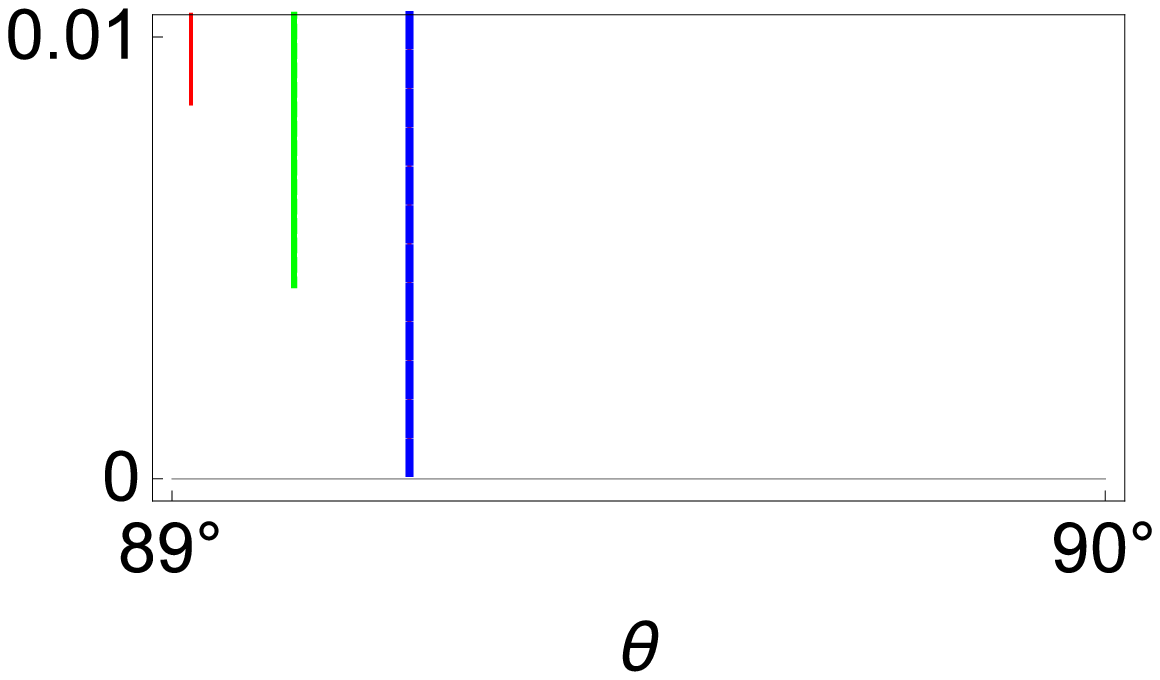}
    \caption{(Color online) Plots of $\left|\textrm{R}^{l}\right|^2$ (thick dashed blue curve), $\left|\textrm{R}^{r}\right|^2$ (thin red curve) and $\left|\textrm{T}-1\right|^2$ (thin dashed green curve) as a function of incidence angle $\theta$. The smaller incidence angle, the better reflectionless and invisible configurations.}
    \label{figpi5}
    \end{center}
    \end{figure}
Finally, to show up how the angle of incidence affects the quantities $\left|\textrm{R}^{l}\right|^2$, $\left|\textrm{R}^{r}\right|^2$ and $\left|\textrm{T}-1\right|^2$, and thus unidirectional reflectionlessness and invisibility phenomena, we refer to Fig.~\ref{figpi5}. We again use the same parameters as in previous cases. We observe that angle range of unidirectional reflectionlessness and invisibility decreases, and frequency of invisibility is lower compared to reflectionlessness. The best angle choices are the ones which are very close to zero.

    \begin{figure}
    \begin{center}
    \includegraphics[scale=.45]{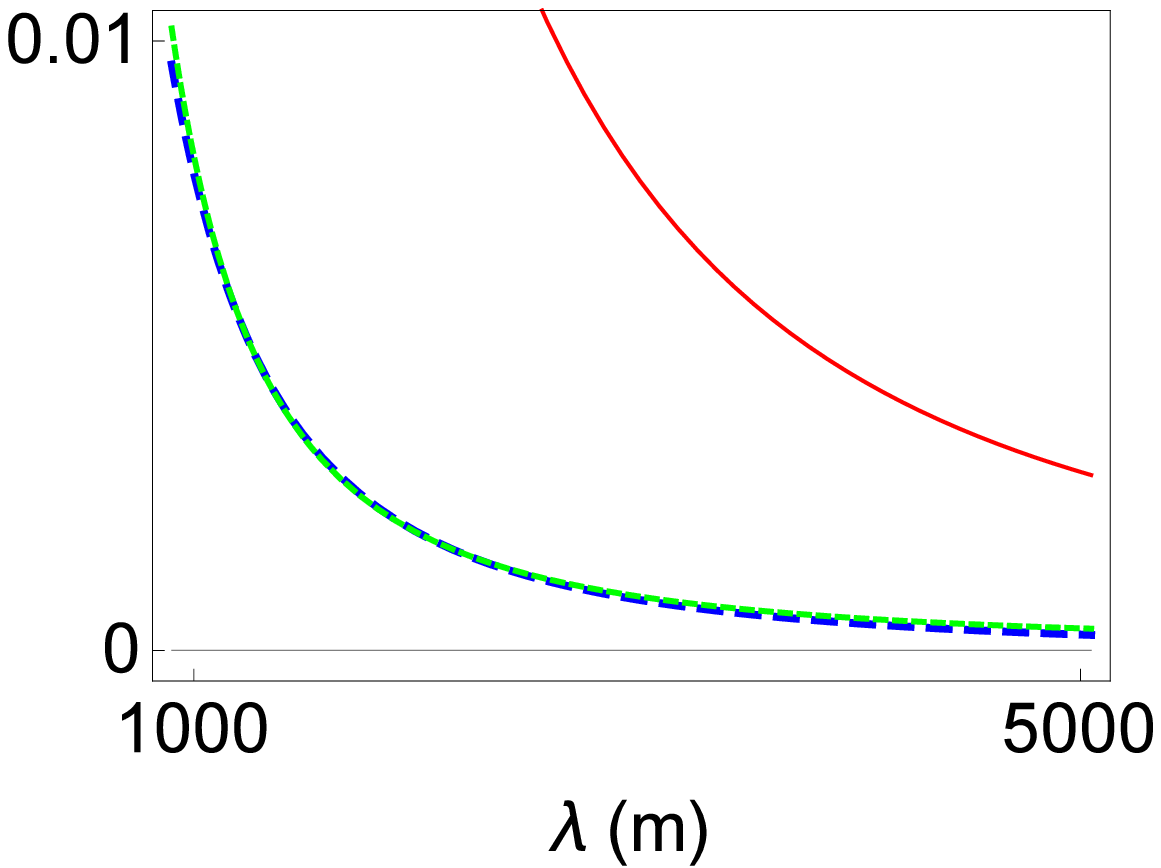}
    \includegraphics[scale=.45]{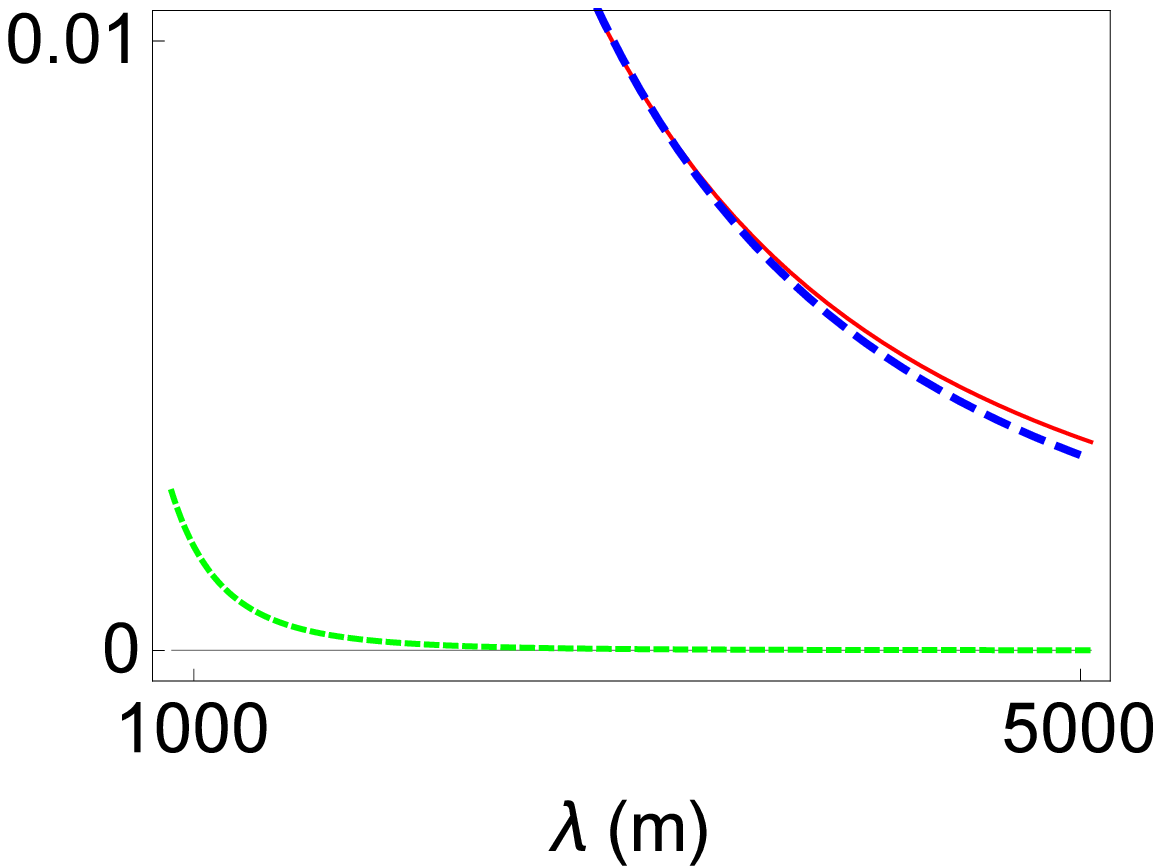}\\
    \includegraphics[scale=.43]{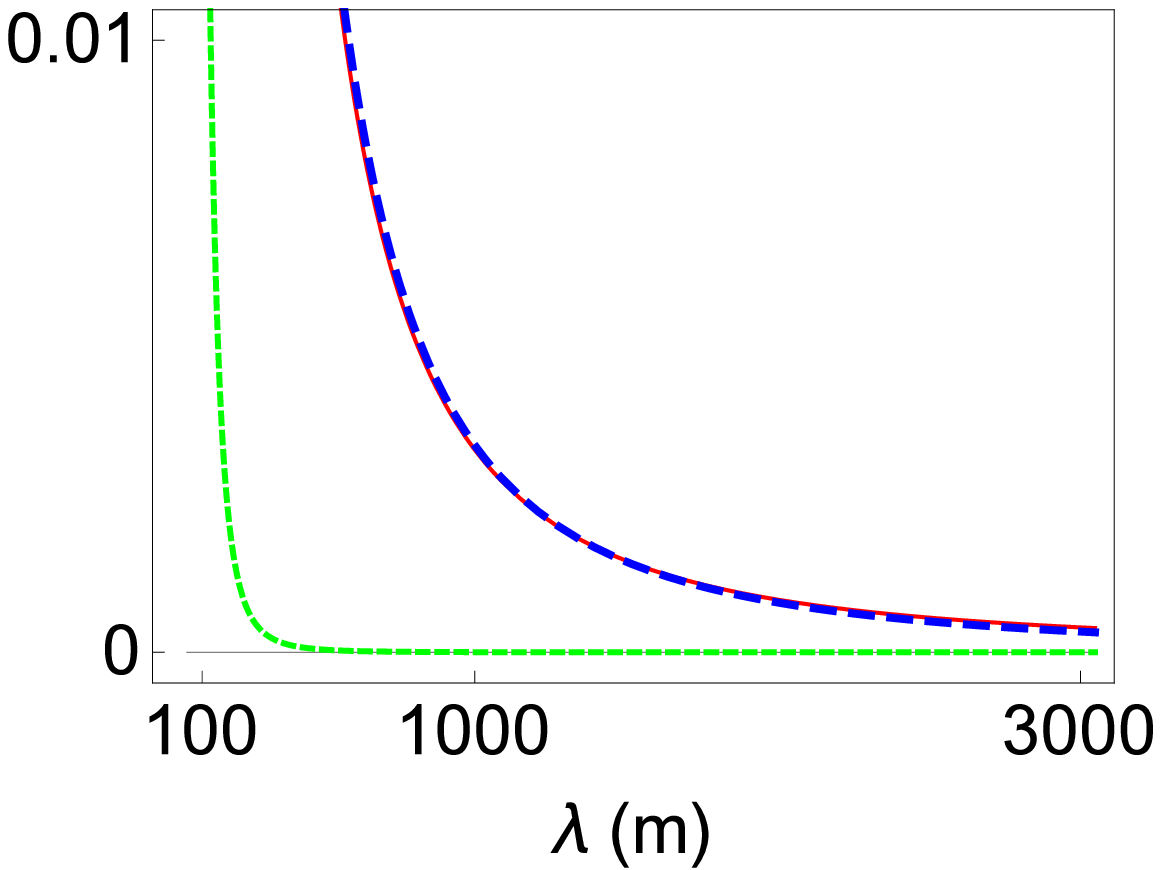}
    \includegraphics[scale=.43]{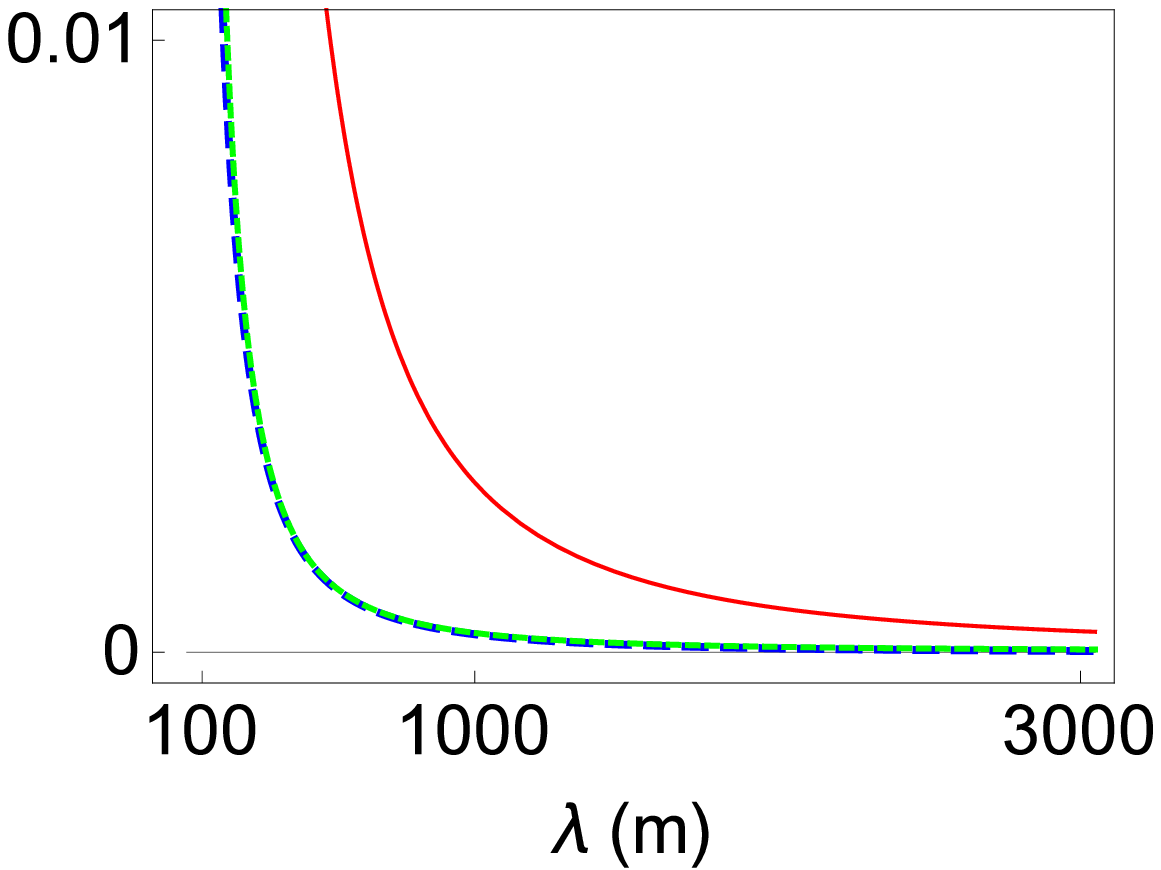}
    \includegraphics[scale=.43]{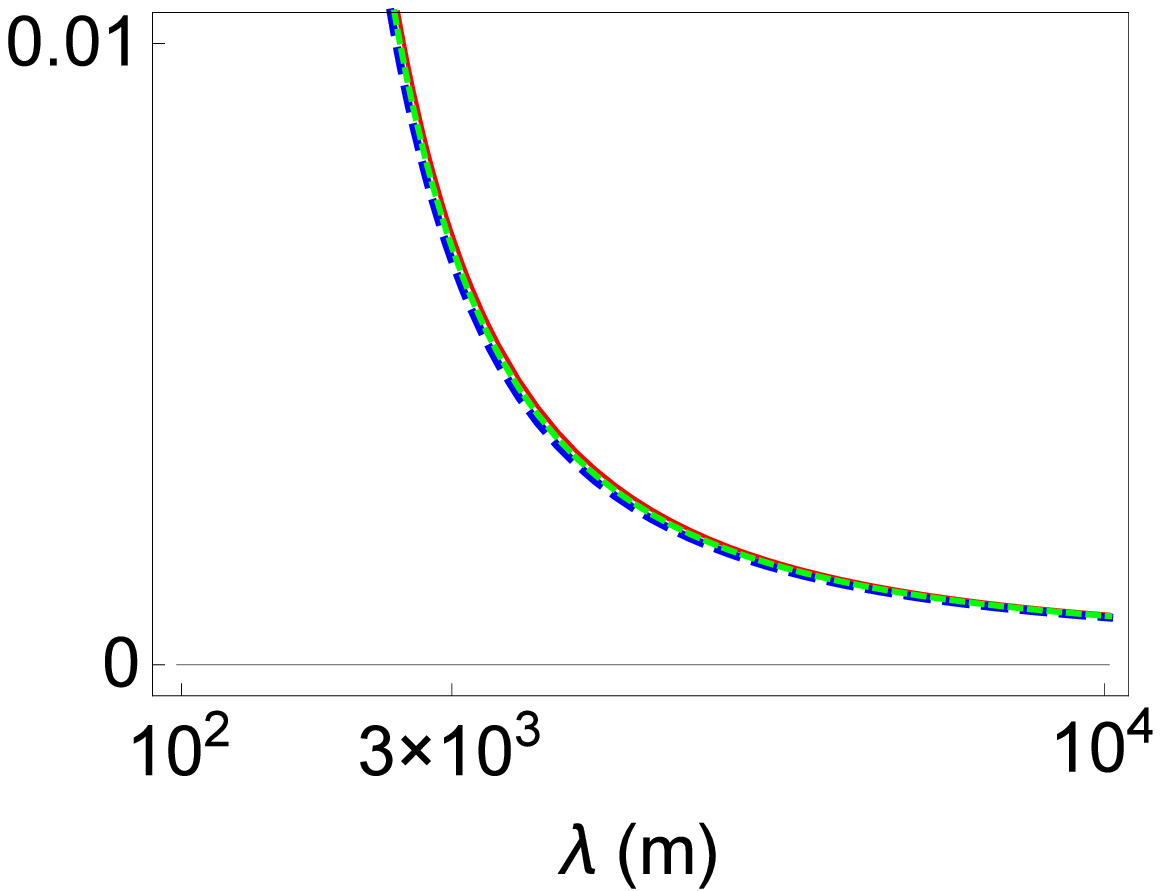}
    \caption{(Color online) Plots of $\left|\textrm{R}^{l}\right|^2$ (thick dashed blue curve), $\left|\textrm{R}^{r}\right|^2$ (thin red curve) and $\left|\textrm{T}-1\right|^2$ (thin dashed green curve) as a function of wavelength $\lambda$ for slab thicknesses $L = 50~\textrm{nm}$ (upper row) and $L = 10~\textrm{nm}$ (lower row) using the slab material as aluminum. As the incidence angle changes, one observes broadband unidirectional reflectionless and invisible configurations.}
    \label{figpi6}
    \end{center}
    \end{figure}
In Fig.~\ref{figpi6}, we reveal the consistency of our findings that parameters of the system has to satisfy in order to generate the desired broadband reflectionless and invisible configurations. For this purpose, we pick out the aluminum as the slab material whose refractive index is $\eta_{Al} = 1.0972$~\cite{alum}, and use two different thickness as $L=50~\textrm{nm}$ (top row figures) and $L=10~\textrm{nm}$ (bottom row figures) to show the effect of thickness. We employ very low temperature and chemical potential values for the graphene as we have explored in Figs. (\ref{figpi3}) and (\ref{figpi4}), $T = 5^{\circ}\textrm{K}$ and $\mu = 5\times 10^{-7}~\textrm{eV}$. We realize that distinct phenomena are observed within $1\%$ of flexibility depending on the incidence angle. First figure on top row gives rise to the left invisibility in broad wavelength range $(900~\textrm{nm}, 3300~\textrm{nm})$, and the right reflectionlessness in the wavelength range $\lambda \geq 2550~\textrm{nm}$ at the angle of incidence $\theta = 62.6^{\circ}$. For the second figure of top row, bidirectional reflectionlessness is obtained at incidence angle $\theta = 40.8^{\circ}$ within the wavelength range $(2800~\textrm{nm}, 3200~\textrm{nm})$. Although we choose a specific incidence angles yielding corresponding phenomena, there is about $1^{\circ}$ angle of extensibility for the same phenomena. For all other angles one can easily find out the left and right reflectionless configurations. As to the second row for the slab thickness $L = 10~\textrm{nm}$, our phenomena in spotlight is observed in a manner that corresponding wavelength ranges widen and cover a broad spectrum even less than visible spectrum. First figure at the bottom row yields a bidirectional reflectionlessness at incidence angle of $\theta = 40.8^{\circ}$ in the wavelength interval $\lambda\geq 550~\textrm{nm}$. When the incidence angle is enlarged to $\theta = 62.8^{\circ}$, we obtain the second figure which gives left invisibility in the wavelength interval $\lambda\geq 165~\textrm{nm}$ and right reflectionlessness in the range $\lambda\geq 500~\textrm{nm}$. Finally, when the incidence angle is $\geq 87$, we obtain bidirectional invisibility within the wavelength interval $\lambda\geq 2370~\textrm{nm}$. At all other angles, left and right reflectionless configurations are obtained. We see that the best invisible and reflectionless configurations are observed once the slab thickness is lowered in nanometric sizes.

\end{document}